\documentclass[aps,pra,showpacs,twocolumn]{revtex4}

\usepackage{amsmath}
\usepackage{amssymb}
\usepackage{mathrsfs}
\usepackage{verbatim}
\usepackage{epsfig}
\usepackage{color}
\usepackage[bookmarks]{hyperref}
\usepackage{graphicx}
\graphicspath{{.}{./EPS/}}

\newcommand* {\bra}[1]{\ensuremath{\langle {#1} |}}
\newcommand* {\ket}[1]{\ensuremath{| {#1} \rangle}}

\newcommand{\cL}{{\mathcal L}}

\newcommand{\cE}{{\mathcal E}}
\newcommand{\cV}{{\mathcal V}}
\newcommand{\Tr}[1]{{\textrm{Tr}}{\left\{#1\right\}}}

\newcommand{\ketbra}[2]{|{#1}\rangle\langle{#2}|}
\newcommand{\braket}[2]{\langle{#1}|{#2}\rangle}

\newcommand{\hbrho}{\hat{\boldsymbol{\rho}}}

\newcommand{\abs}[1]{\ensuremath{\lvert #1 \rvert}}

\newcommand{\expal}{e^{\frac{-\abs{\alpha}^2}{2}}}

\begin{document}
\title{Multiphoton-state-assisted entanglement purification of material qubits}

\author{J\'ozsef Zsolt Bern\'ad}
\email{Zsolt.Bernad@physik.tu-darmstadt.de}
\affiliation{Institut f\"{u}r Angewandte Physik, Technische Universit\"{a}t Darmstadt, D-64289, Germany}
\author{Juan Mauricio Torres}
\affiliation{Institut f\"{u}r Angewandte Physik, Technische Universit\"{a}t Darmstadt, D-64289, Germany}
\author{Ludwig Kunz}
\affiliation{Institut f\"{u}r Angewandte Physik, Technische Universit\"{a}t Darmstadt, D-64289, Germany}
\author{Gernot Alber}
\affiliation{Institut f\"{u}r Angewandte Physik, Technische Universit\"{a}t Darmstadt, D-64289, Germany}

\date{\today}

\begin{abstract}
We propose an entanglement purification scheme based on material qubits and ancillary coherent multiphoton states. We consider a typical QED scenario where
material qubits implemented by two-level atoms fly sequentially through a cavity and interact resonantly with a single-mode of the radiation field. We explore the theoretical
possibilities of realizing a high-fidelity two-qubit quantum operation necessary for the purification protocol with the help of a postselective balanced homodyne 
photodetection. We demonstrate that the obtained probabilistic quantum operation can be used as a bilateral operation in the proposed purification scheme. It is shown that 
the probabilistic nature of this quantum operation is counterbalanced in the last step of the scheme where qubits are not discarded after inadequate qubit measurements. 
As this protocol requires present-day experimental setups and generates high fidelity entangled pairs with high repetition rates, 
it may offer interesting perspectives for applications in quantum information theory.
\end{abstract}

\maketitle

\section{Introduction}

Entanglement purification \cite{Bennett1,Deutsch} is an important protocol which overcomes detrimental effects of noisy channels and generates high fidelity pure entangled states 
from a large number of not-too-low fidelity states. The controlled-NOT gate stays at the core of the protocol and it was experimentally demonstrated 
earlier than the proposal for the entanglement purification \cite{Monroe}. First experimental implementations were done
more than a decade ago using photonic qubits \cite{Zeilinger} and  material qubits \cite{Wineland}. The purification protocol has found application in the proposals for 
quantum repeaters \cite{Sangouard} which enables long distance quantum key distribution \cite{Ekert} and quantum teleportation \cite{Bennett2}. 
The quantum repeater proposed by Ref. \cite{Briegel} has three 
sequentially applied building blocks: in the first step an entanglement is generated between neighboring nodes; in the second step entanglement purification is carried out over the 
ensemble of low-fidelity entangled pairs; in the last step the entanglement swapping procedure transforms the entangled states on the neighboring stations into entangled states 
on the second neighboring stations, thus increasing the distance of shared entanglement. 
There is a specific 
quantum repeater based on hybrid protocols combining continuous and discrete variables \cite{vanLoock1, vanLoock2, vanLoock3}. 
We have already discussed two building blocks of a hybrid quantum repeater scheme 
\cite{Bernad1, Bernad2, Torres} based on coherent multiphoton states and resonant matter-field interactions
which have advantages in the photonic postselection measurements \cite{Bernad1}.
Additionally,  multiphoton coherent states can be produced with high repetition rates and 
they have high transmission rates in the channels connecting the quantum nodes. For example, in long distance quantum key distribution scenarios coherent states 
with both low \cite{Korzh} and high mean photon numbers \cite{Joquet} have already been successfully applied. 
Recently, 
an entanglement purification scheme has been proposed in the context of the hybrid quantum repeater using chains of atoms, optical cavities and far-off 
resonant matter-field interactions \cite{Gonta}. The difficulty in doing this
is due to the long interaction times or large number of photons involved in such a QED scenario. While single-mode fields with high mean photon numbers are not an experimental issue, 
the justification of far-off resonant matter-field interactions requires significant difference between the frequency of the material transition and the frequency of the single-mode field
and this difference has to be further increased with the increase of the mean photon number in the cavity.

In this paper we discuss  entanglement purification schemes which are based on resonant interactions 
between flying material qubits and a single-mode cavity field \cite{Haroche}. At the core of our scheme is the one-atom maser 
which has been experimentally investigated over the last few decades \cite{Rempe}.
Our motivation is to augment our previous work with the missing entanglement purification protocol. Thus we require that the chosen scheme, 
though being not the only possibility to realize an entangling quantum operation \cite{Cirac}, must be compatible with the architecture of a hybrid quantum repeater based 
on coherent multiphoton states and resonant matter-field interactions. We focus on  resonant matter-field interactions between material qubits 
and a single-mode cavity prepared initially in a coherent state. 
The two material qubits fly sequentially through the cavity and interact with the single-mode field resulting in a joint quantum state which after a successful postselective
balanced homodyne photodection yields an entangling two-qubit quantum operation. We demonstrate that this probabilistic quantum operation can replace the controlled-NOT gate
in the purification schemes of Refs. \cite{Bennett1} and \cite{Deutsch}. Furthermore, in our schemes the qubits do not have to be discarded after inadequate qubit measurement results. 
There is a specific Bell diagonal state which is generated in hybrid quantum repeaters and thus being a good example of the purification scheme of Ref. \cite{Deutsch}. 
We discuss the performance of our proposed purification protocols in this specific scenario. Furthermore,  we also investigate the role of the spontaneous decay in the 
material qubits and the damping of the cavity field mode. Thus we present a truly microscopic model of this QED scenario. 

This paper is organized as follows. In Sec. \ref{Model} we introduce the theoretical model. In Sec. \ref{Entmodel} we determine the form of the two-qubit quantum operation 
which is generated by a postselective balanced homodyne photodetection. Numerical results are presented for the success probability of obtaining this quantum operation. 
These results are employed in Sec. \ref{Purify} to realize entanglement purification. In Sec. \ref{loss} we study the role of spontaneous decay and cavity losses and their 
effect on the purification schemes. Details of the relevant photon states of the unitary model are collected in the Appendix.

\section{Model}
\label{Model}

\begin{figure}[t]
 \includegraphics[width=.35\textwidth]{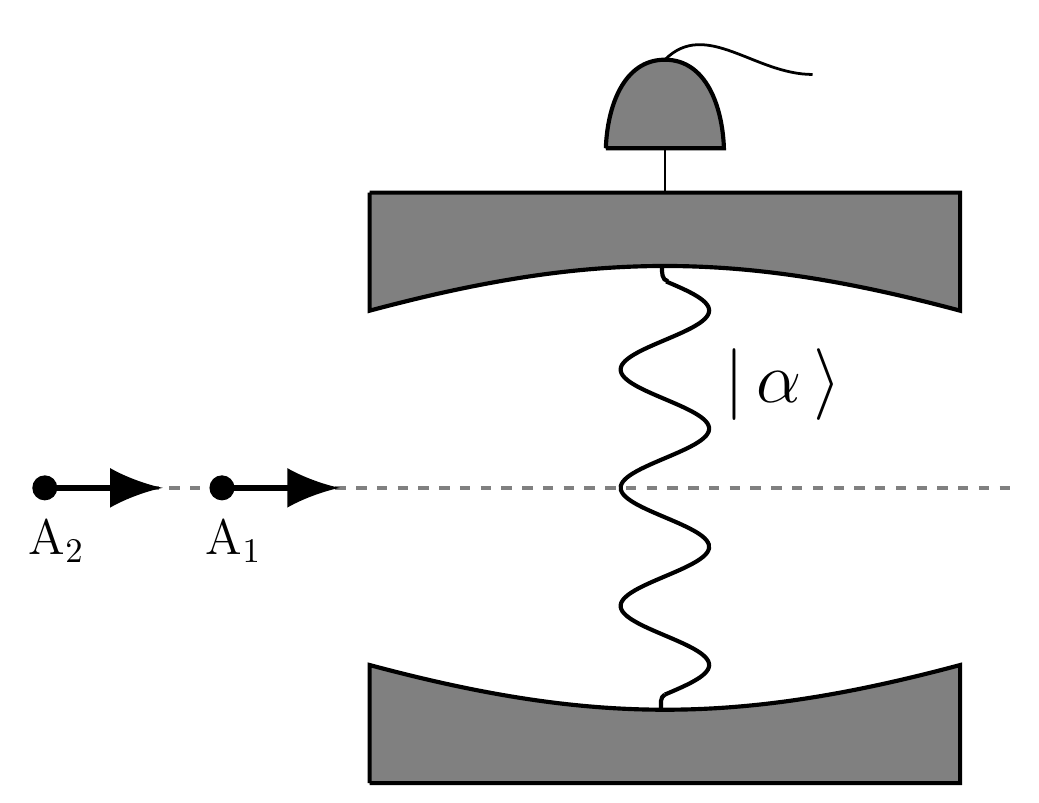}
  \caption{A cavity-QED setup for a probabilistic two-qubit quantum operation. Two qubits $A_1$ and $A_2$ fly sequentially through a cavity 
  and they interact resonantly with a single-mode field. The field is initially prepared in a coherent state $\ket{\alpha}$. 
  After both qubits passed through the cavity, the field state is postselected by a balanced homodyne detection which is depicted as a detector outside the cavity. 
  Provided that we are successful the resulting two-qubit quantum operation is applied in the entanglement purification schemes in Sec. \ref{Purify}.}
   \label{fig:setup}
\end{figure}

In this section we discuss a QED model consisting of a single-mode cavity in which two atoms, implementing the material qubits, are injected sequentially such that at most
one atom at a time is present inside the cavity. The field inside the cavity is prepared initially in a coherent state and after both interactions the state of the field
gets correlated with the state of the qubits. This scenario, illustrated in Fig.~\ref{fig:setup}, is motivated by the progress in atom-cavity implementations, 
whereas with the help of cutting-edge technology all the relevant parameters which justify our setup are well under control \cite{Haroche,Haroche2}. 
We present the solution to this model and discuss its properties with the help of the coherent state approximation \cite{Gea-Banacloche}.

Let us consider two qubits $A_1$ and $A_2$ with ground states $\ket{0}^\ell$ and excited states 
$\ket{1}^\ell$ ($\ell \in\{A_1,A_2\}$). These qubits pass through a cavity in sequence and interact with
a single-mode radiation field which is in resonance with the qubit's transition frequency.
This corresponds to the well-known resonant Jaynes-Cummings-Paul interaction \cite{Jaynes,Paul}. 
Due to the resonant condition we are going to work in a 
time independent interaction picture with respect to the free energy of the cavity and the two qubits. 
In the dipole and rotating-wave approximation 
the Hamiltonian accounting for the dynamics of qubits and field is given by ($\hbar=1$)
\begin{align}
  \hat H^\ell=g \left(\hat{a} \hat{\sigma}^\ell_+ +  \hat{a}^{\dagger} \hat{\sigma}^\ell_- \right),
  \quad \ell \in \{A_1,A_2\}.
  \label{}
\end{align}
We have considered the raising and lowering operators
$\hat{\sigma}^\ell_+=\ket{1} \bra{0}^\ell$ and $\hat{\sigma}^\ell_-=\ket{0}\bra{1}^\ell$, 
and  
the vacuum Rabi splitting $2g$ for the dipole transition. Furthermore,
$\hat{a}$ ($\hat{a}^\dagger$) is the destruction (creation) operator of the field mode. 

We are interested in the situation where there are no initial correlations between the field and the qubits. 
Therefore, we choose an initial state of the form
\begin{equation}
\ket{\Psi_0}= \left (c_{00} \ket{00} + c_{01} \ket{01} + c_{10} \ket{10} + c_{11} \ket{11} \right) \ket{\alpha},
\label{initial}
\end{equation}
with the qubits set in an arbitrary state in the basis 
$\ket{ij}=\ket{i}^{A_1}\ket j^{A_2}$ ($i,j \in \{0,1\}$), 
%$|c_{00}|^2+|c_{01}|^2+|c_{10}|^2+|c_{11}|^2=1$ 
and the field is in a coherent state
\begin{equation}
  \label{chst}
 \ket{\alpha}=\sum_{n=0}^\infty 
  e^{-\frac{|\alpha|^2}{2}}
 \frac{\alpha^n}{\sqrt{n!}}
 \ket n,
  \quad\alpha=\sqrt{\bar n}\,e^{i\phi}
\end{equation}
written in terms of the photon-number states $\ket{n}$ ($n\in {\mathbb N}_0$)
and with the phase $\phi$.
As stated before, we are interested in the most simple scenario where the two qubits 
interact independently and sequentially with the field. Therefore, the evolution 
operator $\hat{U}(t)$ can be written as a product of separate evolution operators and the temporal
state vector can be evaluated as
\begin{align}
  \ket{\Psi(\tau)}=
  \hat U(\tau)\ket{\Psi_0},\quad
 \hat U(\tau)=  e^{-i \hat H^{A_2}\tau} e^{-i \hat H^{A_1}\tau},
  \label{evolution}
\end{align}
where we considered equal interaction times.

Solving the state vector is not a complicated task as it is based on the well known solutions 
of the resonant Jaynes-Cummings-Paul model (see for example \cite{Schleich}). The result
is a time dependent quantum state $\ket{\Psi(\tau)}$ of the tripartite system that 
can be expressed in the following form
\begin{eqnarray}
  \ket{\Psi(\tau)}&=&\ket{00}\ket{g_{00}(\tau)}
  +\ket{01}\ket{g_{01}(\tau)}
  +\ket{10}\ket{g_{10}(\tau)} \nonumber \\
  &+&\ket{11}\ket{g_{10}(\tau)},
  \label{psi}
\end{eqnarray}
where the unnormalized field states $\ket{g_{ij}(\tau)}$ are presented in Appendix \ref{App}. 

%%%%%%%%%%%%%%%%%%%%%%%%%%%%%%%%%%%%
%%%%%%%%%%%%%%%%%%%%%%%%%%%%%%%%%%%%%%%

In order to obtain a better understanding of the field states we concentrate on the case of 
large mean photon number $\bar n \gg 1$ and interaction times $\tau$
such that the Rabi frequency $g \sqrt{n}$ can be linearised around $\bar n$. 
This procedure can be justified for short interaction times $\tau$ that fulfill the condition
$g\tau\ll \sqrt{\bar n}$.
This corresponds to a time scale well below the well-known revival phenomena of the population inversion in the 
Jaynes-Cummings-Paul model \cite{Cummings,Eberly}.
Thus one can find that the state of Eq. \eqref{psi} can be approximated by
\begin{equation}
\ket{\Psi(\tau)}\approx
\ket{\psi_-} 
\ket{\alpha e^{-i\frac{g}{\sqrt{\bar n}}\tau}}  
+ \ket{\psi_+}  
\ket{\alpha e^{i\frac{g}{\sqrt{\bar n}}\tau}}  
+
\ket{\psi_\star} \ket{\alpha}
 \label{psi_mpn}
\end{equation}
with the two-qubit unnormalized states
\begin{align}
  \ket{\psi_\star}&=
  \frac{c_{01}-c_{10}}{\sqrt2}\ket{\Psi^-}
  +
  \frac{c_{00}e^{i\phi}-c_{11}e^{-i\phi}}{\sqrt2}\ket{\Phi^-_{\phi}},
  \label{Malpha}
  \\
  \ket{\psi_\pm}&=
  \frac{c_{00}e^{i\phi} + c_{11} e^{-i \phi} \mp c_{01}\mp c_{10}
}{2
e^{\mp ig\sqrt{\bar n}\tau}}
\frac{\ket{\Phi^+_{\phi}}\mp\ket{\Psi^+} }{\sqrt2}
  \label{Mmp}
\end{align}
which have been written in terms of the  Bell states
\begin{align}
  \ket{\Psi^\pm}&=\frac{1}{\sqrt2}\left(\ket{01}\pm\ket{10}\right),
\\
  \ket{\Phi^\pm_\phi}&=\frac{1}{\sqrt2}\left(e^{-i\phi}\ket{00}\pm e^{i\phi}\ket{11}\right),
  \quad \ket{\Phi^\pm}=\ket{\Phi^\pm_0}.
  \label{Bellstates}
\end{align}
One can note that the state in Eq. \eqref{psi_mpn} involves only three coherent states: 
two of them are $\ket{\alpha e^{\pm ig\tau/\sqrt n}}$ that rotate with frequencies of opposite sign and
a third one $\ket{\alpha}$ which corresponds to the initial coherent sate. The approximation in Eq.
\eqref{psi_mpn} makes evident that a postselective field measurement can be used to prepare an entangled two-qubit state.
Of course the simplest non-trivial situation is when the three coherent states are nearly orthogonal. For this 
purpose we consider the overlaps
between $\ket{\alpha}$, $\ket{\alpha e^{\pm ig \tau/\sqrt n}}$ that yield
\begin{equation}
  \langle \alpha \ket{\alpha e^{\pm i\frac{g}{\sqrt{\bar n}}\tau}}=  
  \exp{\left[-\bar n \left(1-e^{\pm i\frac{g}{\sqrt{\bar n}}\tau}\right)\right]}
  \approx e^{- \frac{g^2\tau^2}{2}}.
 \label{overlap}
\end{equation}
The last approximation holds for $g\tau \ll \sqrt{\bar n}$ and shows that the overlap
nearly vanishes for interaction times $g\tau>\sqrt2$. 
It can be shown that the overlap between the other two states vanishes faster in time. Therefore we
consider interaction times that fulfill the condition 
\begin{equation}
  \sqrt 2<g\tau\ll\sqrt{\bar n}.
 \label{taucondition}
\end{equation}
We emphasize that the first inequality is to ensure orthogonal field states, while
the second inequality sets a bound in time where the coherent state approximation is valid.
We close this section by pointing out an interesting fact that a similar result to 
Eq. \eqref{psi_mpn} can be obtained by choosing a setup where the two qubits
interact simultaneously with a single-mode field for a time $\tau$. In our previous works
\cite{Torres,Torres2} we have shown that in the coherent state approximation the two-atom Tavis-Cummings model
results in a solution where the two-qubit state $\ket{\psi_\star}$ is paired up with $\ket\alpha$.

\section{Entangling quantum operation}
\label{Entmodel}
\subsection{Postselection by projection onto $\ket\alpha$}

Our subsequent investigation is to determine a field measurement which is capable to realize conditionally an 
entangling two-qubit quantum operation.  
Eq. \eqref{overlap} shows that the overlaps between the coherent states approximately vanish for interaction times 
$g \tau>\sqrt2$. Thus a postselective measurement on the field states has the 
possibility to generate three two-qubit quantum operations which used on the initial state in Eq. \eqref{initial} 
result in the states of Eqs. \eqref{Malpha} and  \eqref{Mmp}. However, only the two-qubit state
in Eq. \eqref{Malpha} is a good candidate for an entanglement purification scheme. The reason is that 
the states $\ket{\psi_{\pm}}$ are separable states. Only the state $\ket{\psi_\star}$ has the potential 
to be entangled. 
In order to postselect the state $\ket{\psi_\star}$ one has to implement the following quantum 
operation for any initial two-qubit state $\ket{\psi}$
\begin{align}
  \hat W_1{(\phi,\bar n)}\ket\psi
  =\bra{\alpha}\hat U(\tau)\ket{\psi}\ket\alpha
  \approx 
  \ket{\psi_\star}.
  \label{W1}
\end{align}
The operation is performed by
first letting the qubits interact with the field, as depicted in Fig. \ref{fig:setup}.
This is described by the evolution $\hat U(\tau)\ket{\psi}\ket\alpha$, with the evolution operator
in Eq. \eqref{evolution}. After the interaction, the state of the field is projected onto the coherent state $\ket\alpha$.  
By appropriate values of $\bar n=|\alpha|^2$ and $\tau$ (see Eq. \eqref{taucondition})
this operation approaches the quantum operation
$\hat W_1{(\phi,\bar n)}\to\hat M_\phi$ that can be 
represented as the sum of two projectors on Bell states:
%in the standard basis as the following matrix:
\begin{equation}
  \hat{M}_\phi=\ketbra{\Psi^-}{\Psi^-}+\ketbra{\Phi^-_\phi}{\Phi^-_\phi}, \quad \hat M=\hat M_0.
% \frac{1}{2}
% \begin{bmatrix} 1 & 0 & 0 & -e^{- 2i \phi} \\ 0 & 1 & -1 & 0 \\ 0 & -1 & 1 & 0 \\ -e^{2i \phi} & 0 & 0 & 1 \end{bmatrix},
%   \quad
%   \hat M=\hat M_0.
 \label{Qgate}
\end{equation}
In particular, its action on the atomic states of the standard basis can be listed as 
\begin{eqnarray}
  \hat M_\phi\ket{01}=\tfrac{1}{\sqrt2}\ket{\Psi^-},\quad 
  \hat M_\phi\ket{10}=-\tfrac{1}{\sqrt2}\ket{\Psi^-}, 
 \nonumber \\
  \hat M_\phi\ket{00}=\tfrac{e^{i\phi}}{\sqrt2}\ket{\Phi^-_\phi},\quad 
  \hat M_\phi\ket{11}=-\tfrac{e^{-i\phi}}{\sqrt2}\ket{\Phi^-_\phi}. 
  \nonumber 
\end{eqnarray}
This entanglement generating property of $\hat M_\phi$ allows us to use it as a bilateral operation in entanglement purification schemes of Refs. \cite{Bennett1} 
and \cite{Deutsch} as we will show in Sec. \ref{Purify}. A practical question is how to realize the postselective measurement of the field. In the next subsection
we investigate this issue by means of balanced homodyne photodetection \cite{Lvovsky}.

\subsection{Postselection by balanced homodyne photodetection}
%%%%%%%%%%%%%%%%%%%%%%%%%%%%%%%%%%%%
\begin{figure*}[ht!]
 \begin{center}
 \includegraphics[width=.51\textwidth]{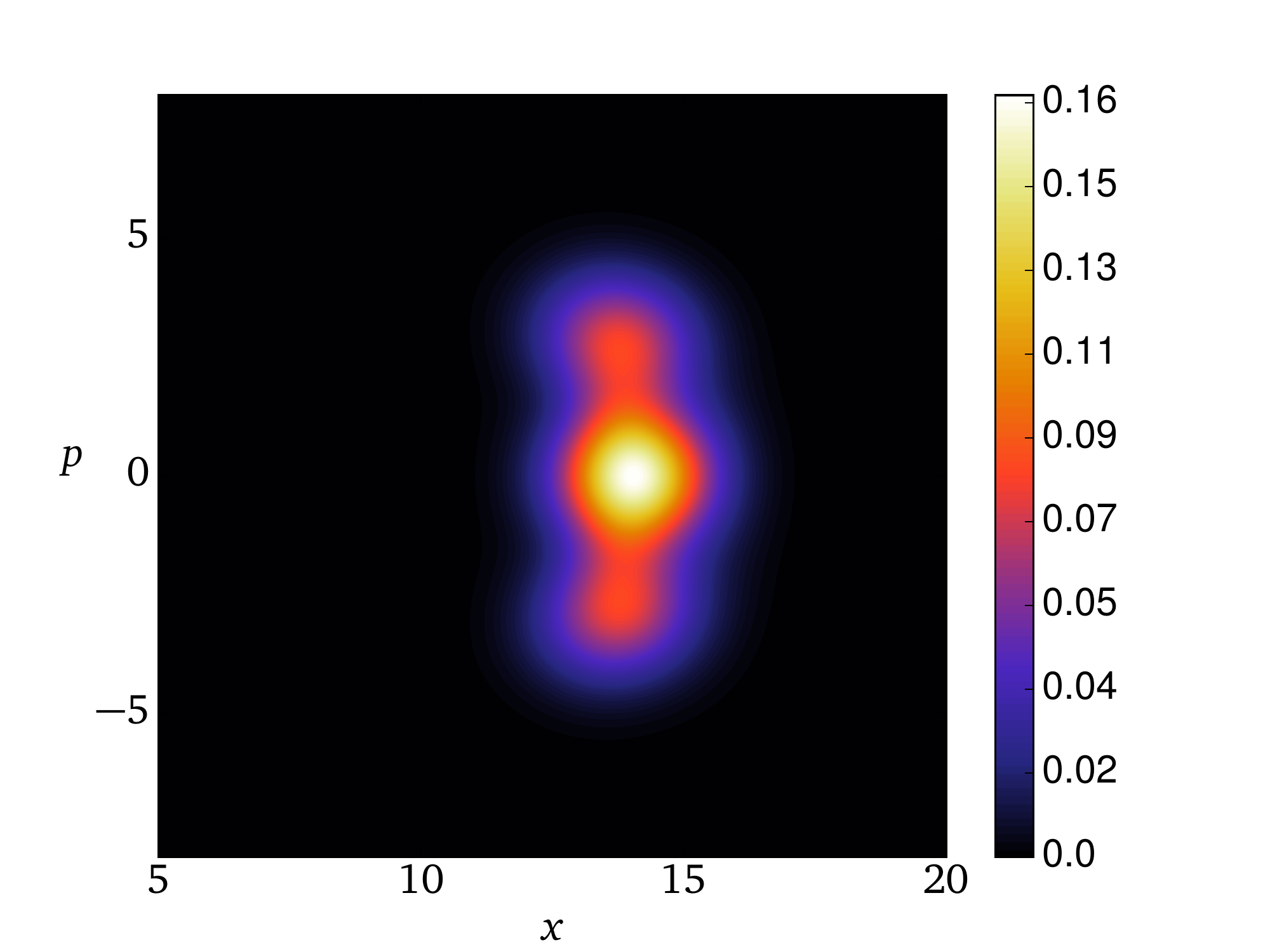}
 \includegraphics[width=.48\textwidth]{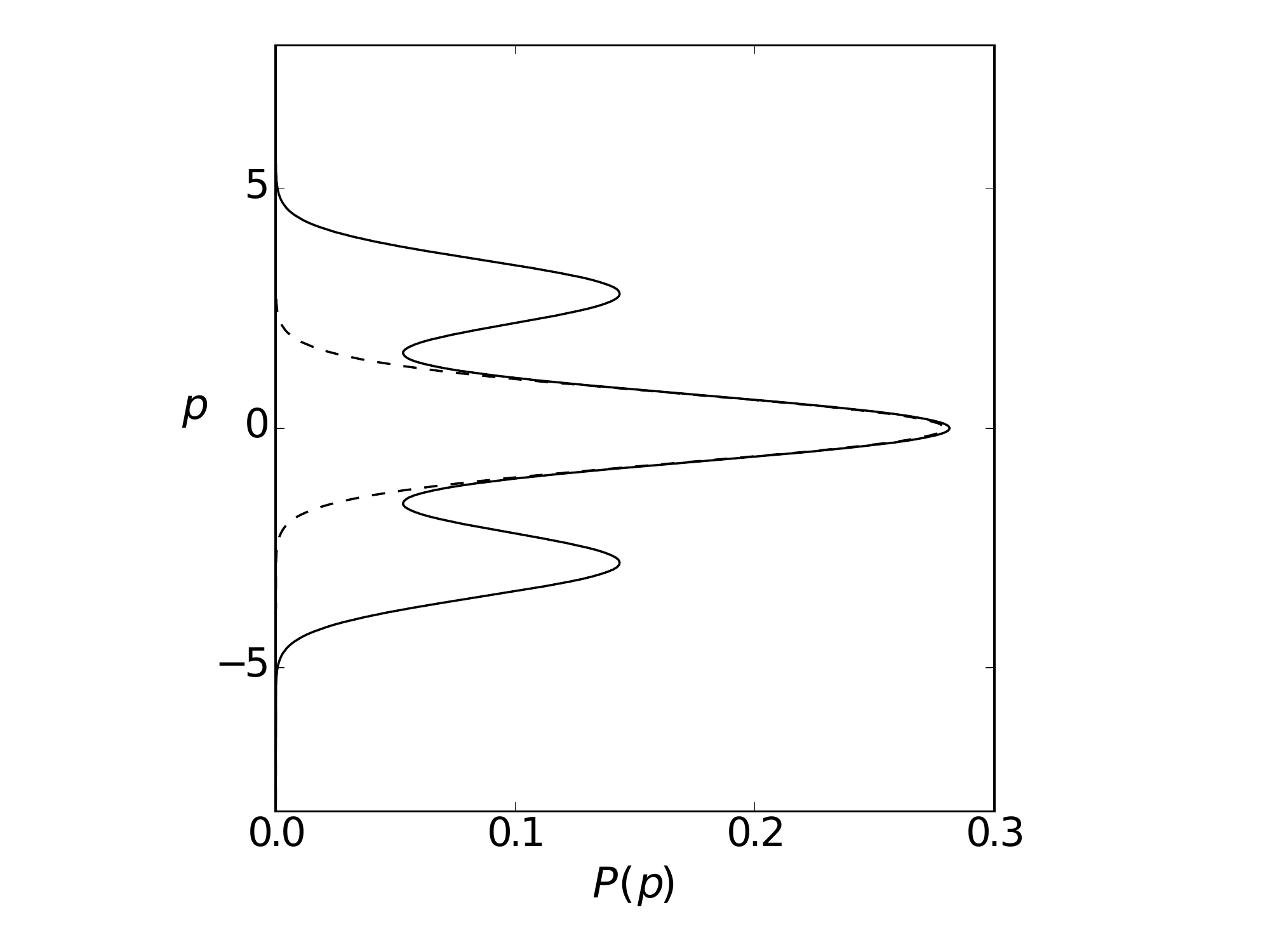}
\caption{
Left panel: Husimi Q-function of the field state defined in Eq. \eqref{Husimi}
after the interaction between the cavity and the qubits as depicted in Fig. 
\ref{fig:setup}. Right panel: The corresponding probability distribution $P(p)$ for the $p$ quadrature
defined in Eq. \eqref{probhomod} in full line  and the weighted $p$ quadrature distribution of the initial coherent state $\ket{\alpha}$ as a reference in dashed line.  
The interaction time is $\tau=2/g$. The initial tripartite state in Eq. \eqref{initial} is considered to be $\ket{00} \ket{\alpha}$ 
with $\alpha =10$.
  }
   \label{fig:QA}
 \end{center}
\end{figure*}
%%%%%%%%%%%%%%%%%%%%%%%%%%%%%%%%%%%%%%%

In the following we return to our exact calculations in Eq. \eqref{psi} and show that this quantum 
operation can be probabilistically implemented with fidelity close to unity by  measuring the state of the field with a
balanced homodyne photodetection. First let us study the evolution of the field in  phase space
with dimensionless coordinates $x$ and $p$. We use for this purpose the Husimi Q-function, defined as
\begin{equation}
  Q(\beta,\tau) = \frac{1}{\pi} \bra{\beta} \hat{\rho}^F(\tau)\ket{\beta}, \quad \beta=\frac{x+i\,p}{\sqrt2},
  \label{Husimi}
\end{equation}
where we have introduced  the reduced density matrix of the field state
$\hat{\rho}^F(\tau) = \mathrm{Tr}_{A_1,A_2}\left\{\ket{\Psi(\tau)}\bra{\Psi(\tau)}\right\}$ 
with $\mathrm{Tr}_{A_1,A_2}$ being the partial trace over the qubits. 
Figure \ref{fig:QA} shows the $Q$ function of the field after the interactions with the two qubits.
The initial field state is characterized by $\alpha=10$, i.e., $\phi=0$.
Although the results can be extended to arbitrary values of $\phi$, for
the sake of simplicity here and in the rest of the paper we consider $\phi=0$. 
It can be noted that the $Q$ function is composed of three spots, each of which corresponds to a coherent state in Eq. \eqref{psi_mpn}.
During the first interaction the initial coherent state
splits into  two spots that evolve with frequencies of opposite sign.
When qubit $A_2$ interacts with the field emerged after the interaction with qubit $A_1$, 
both spots split up again. Due to the fact that the interaction time for both of the qubits is equal, 
the spots moving backwards meet again at the initial position. Furthermore,
the state at the initial position is close to a coherent state while the two other spots are deformed due to the nonlinear dependence
of the Rabi frequencies on the photon-number. It is an interesting feature that the initial coherent state is almost restored and this
makes the central contribution to the field state an attractive candidate
to be measured. Provided that we are successful in this measurement we generate the two-qubit  quantum operation in Eq. \eqref{Qgate}. 

In the next step we focus on the postselective field measurement.
We briefly recapitulate the basic features which lead to a quadrature 
measurement of the field with the help of a balanced homodyne measurement 
\cite{Lvovsky,Torres}.
The field state, subject to detection, is superposed with a strong local coherent state, i.e. high mean number of photons, at a $50\%$ reflecting beam splitter, and the modes 
emerging from the beam splitter are measured measured by two photodetectors. We consider in our scheme ideal photodetectors. The measured signal is the difference of photon numbers
of the two photodetectors. Dividing the measured signal by the square root of two times 
the local coherent state's mean photon number results in a signal, which corresponds to a projective measurement of a quadrature operator $\ket{x_\theta}\bra{x_\theta}$
on the field state. The eigenvalue equation of the quadrature $\ket{x_\theta}$ reads
\begin{equation}
\hat x_\theta\ket{x_\theta}=
\frac{1}{\sqrt{2}}\left(\hat{a} e^{-i\theta}+\hat{a}^\dagger e^{i\theta}\right)\ket{x_\theta}=x_\theta \ket{x_\theta},
\label{quadrature}
\end{equation}
where $\theta$ is the phase of the local oscillator. $\hat{a}$ and $\hat{a}^\dagger$ are the annihilation and creation operators 
of the single-mode field to be measured. Here, we assume that the emerged field state in the cavity can be perfectly transferred to this single-mode field. Due to the phase space structure seen in Fig.~\ref{fig:QA} it is reasonable to 
select the phase of the local oscillator to be $\theta=\frac{\pi}{2}$, i.e., the coordinate $p=x_{\pi/2}$. The reason is 
that the field contribution paired with the two-qubit quantum operation is the farthest from the other field contributions 
in this particular quadrature measurement. 
We remark that the results can be extended to arbitrary 
$\phi\neq 0$ by shifting the phase of the quadrature to be measured, i.e., $x_{\phi+\pi/2}$.

In order to postselect the two-qubit state $\ket{\psi_\star}$, one requires to project the field state with the projector
$\ketbra{p}{p}$ restricted to the interval $p\in [-2,2]$. This ensures that the measurement is selecting only the 
middle contribution in phase space that corresponds to the coherent state 
$\ket{\alpha}$ and also the highest probability of postselecting the two-qubit state $\ket{\psi_\star}$. In this case the postselected 
two-qubit quantum operation takes the form
\begin{align}
  \hat W_2{(p,\bar n)}\ket\psi
  &=\bra{p}\hat U(\tau)\ket{\psi}\ket\alpha
  \approx\ket{\psi_\star},
  \nonumber\\
  p&\in[-2,2],\quad \alpha=\sqrt{\bar n}.
  \label{W2}
\end{align}
The probability for such an event is given by
\begin{align}
 P_H=  \int_{-2}^2P(p)dp,\quad
P(p)=
\Tr{\ketbra{p}{p}\,\ketbra{\Psi(\tau)}{\Psi(\tau)}},
\label{probhomod}
\end{align}
which is obtained by integrating the probability distribution of the field $P(p)$ in the $p$ quadrature. In the limit of high mean photon numbers, 
this can be approximated by integrating the function 
\begin{equation}
  P_H\approx \int_{-2}^2 \frac{|\braket{p}{\alpha}|^2}{|\braket{\psi_\star}{\psi_\star}|^2}dp
  =\frac{{\rm erf}(2)}{|\braket{\psi_\star}{\psi_\star}|^2}.
  \label{}
\end{equation}
with the error function ${\rm erf}(2)=0.995322$ \cite{AS}.  For large mean photon numbers $\bar n$ and with the interaction time fulfilling 
condition \eqref{taucondition}  the postselected  two-qubit quantum operation approaches the quantum operation $\hat M$ in Eq. \eqref{Qgate}.

In the right panel of Fig. \ref{fig:QA} we have plotted in full line the distribution 
$P(p)$ rotated $90$ degrees clockwise to have a better comparison with the $Q$ function in the left panel. We have also plotted by a dashed line the distribution 
$|\braket{p}{\alpha}|^2/|\braket{\psi_\star}{\psi_\star}|^2$ to compare with $P(p)$ and
show the difference between the coherent state $\ket{\alpha}$ and the field state $\hat{\rho}^F(\tau)$ emerged after the matter-field interactions. In the case of the 
coherent state the integration over all relevant  quadrature values $p \in [-2,2]$ results in an almost perfect projection onto the coherent state. However, in the case of
the field state $\hat{\rho}^F(\tau)$ this projection is only achieved for certain interaction times and large mean photon numbers.

%%%%%%%%%%%%%%%%%%%%%%%%%%%%%%%%%%%%%%%
\begin{figure}[ht!]
 \begin{center}
 \includegraphics[width=.45\textwidth]{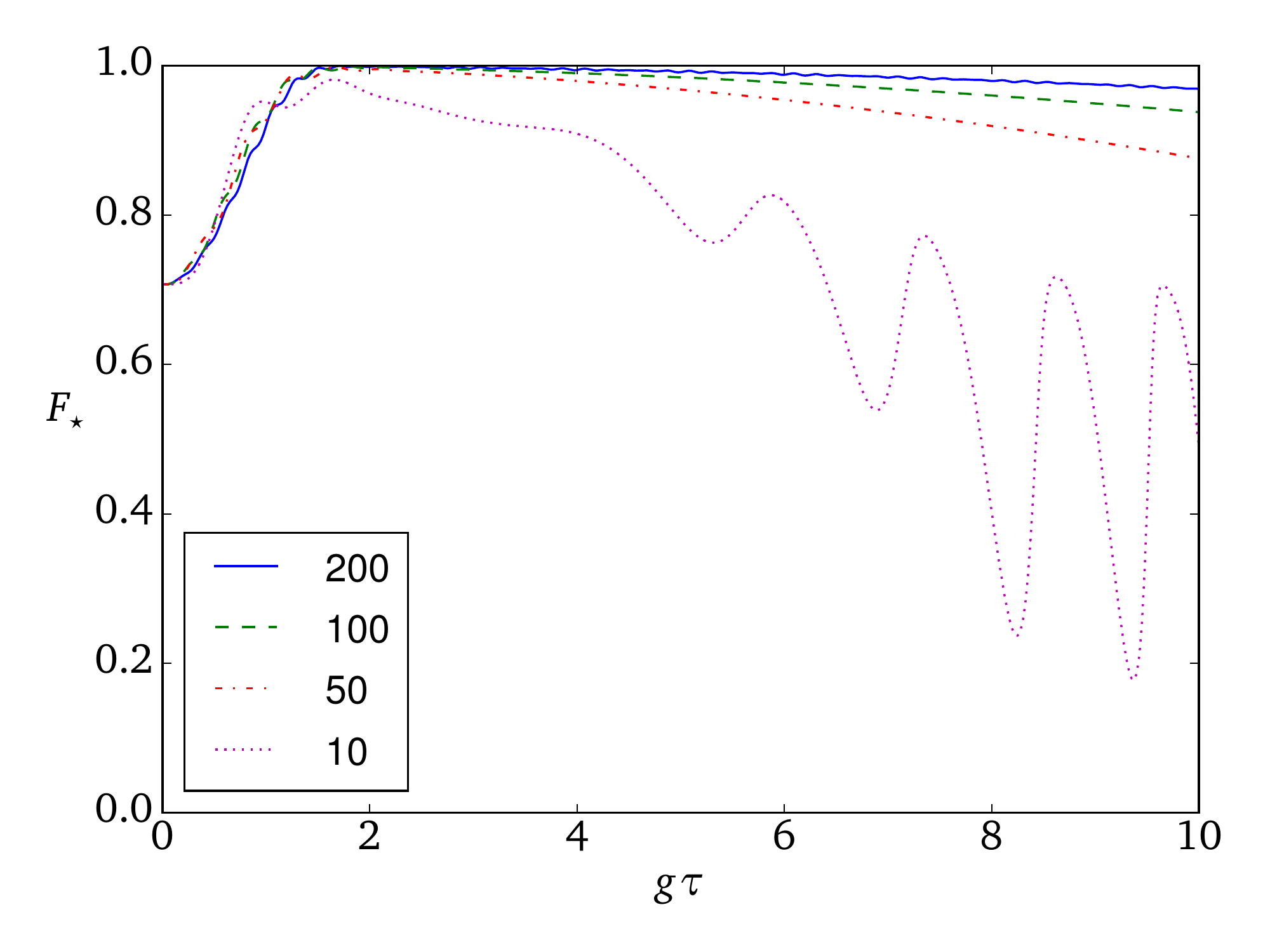}
  \caption{The fidelity $F_\star$ 
  of the two-qubit state in Eq. \eqref{W2} with respect to $\ket{\psi_\star}$ after a successful projective measurement on the quadrature 
  $\ket{p}$, with $p=0$.   The initial state of the two qubits is set to $\ket{00}$ and
 the initial coherent state was taken with real $\alpha=\sqrt{\bar n}$.  Four curves are presented for different values of the mean photon
number $\bar n \in \{10,50,100,200\}$ as described in the legend.}
   \label{fig:fid1}
 \end{center}
\end{figure}
%%%%%%%%%%%%%%%%%%%%%%%%%%%%%%%%%%%%%%%

In order to see how well the state $\ket{\psi_\star}$ can be generated, we consider the fidelity 
\begin{equation}
F_\star = |\bra{\psi_\star} \hat W_2{(p,\bar n)}\ket{\psi}|^2
  \label{fstar}
\end{equation}
after a successful projective measurement on the quadrature $\ket{p}$.  
Figure \ref{fig:fid1} shows the fidelity $F_\star$ as a function of the interaction 
time $\tau$ and for  different values of mean photon number $\bar n$. 
The quadrature measurement was taken always
at the middle of the distribution $p=0$ and qubits were considered initially in the state $\ket{00}$, i.e.,  
both in the ground state. The fidelity increases as a function of time until it reaches it maximum value
around $\tau g=2$. This is the time required for the coherent states of the field to be distinguishable.
Afterwards the fidelity drops down again as the coherent state approximation breaks down with increasing
time. However this decrease in fidelity is slower for larger values of the mean photon number $\bar n$,
in agreement with the limits of the interaction time given in Eq. \eqref{taucondition}.

\section{Entanglement purification}
\label{Purify}

In this section it is demonstrated how the two-qubit quantum operation in Eq. \eqref{Qgate} can be used for implementing entanglement purification schemes. 
The basic idea of entanglement purification is to increase the degree of entanglement of
a qubit pair %$\hat \rho^{A_1,B_1}$ 
at the expense of another qubit pair. %$\hat \rho^{A_2,B_2}$.
Therefore, the protocol can be assumed to start with 
%a four qustateThe initial state can be written as
%to start with 
a product state of two entangled qubit pairs
\begin{equation}
\hbrho=\hat{\rho}^{A_1,B_1} \otimes \hat\rho^{A_2,B_2},
\label{InitPur}
\end{equation}
where $A$ and $B$ are two spatially separated quantum systems. 
The task has to be accomplished by applying local quantum operations and measurements on sides $A$ and $B$ separately.
The measurement procedure leads to the destruction of one of the pairs, say $\hat\rho^{A_2,B_2}$. 
The final result is a qubit pair $\hat \rho'^{A_1,B_1}$ with
a higher fidelity with respect to a maximally entangled sate, typically chosen to be the  Bell state $\ket{\Psi^-}$.
Provided one has a large number of qubit pairs, the iteration of the protocol leads to the distillation of a maximally
entangled state. In the following, we discuss two of the most well-known  protocols \cite{Bennett1,Deutsch} and
present alternative versions using the quantum operation of Sec. \ref{Entmodel}.

{\it Scheme 1.} The first method presented here is based on the pioneering work of Ref. \cite{Bennett1} 
where the entanglement purification protocol distills the entangled state $\ket{\Psi^-}$ 
from a large ensemble of states $\hat{\rho}$ with the property $\bra{\Psi^-} \hat{\rho} \ket{\Psi^-}>\frac{1}{2}$. 
The protocol for two qubit pairs can be summarized in five steps:\\ 
({\bf B1}) Transform both $\hat{\rho}$ into the Werner form. \\
({\bf B2}) Apply $\hat{\sigma}_y^{A_1}$ and $\hat{\sigma}_y^{A_2}$  (Pauli spin operators).\\
%(B3) Two local controlled-NOT gates are performed on the pairs. \\
({\bf B3}) Perform the bilateral operation  $\hat U_{\rm CNOT}^{A_1\rightarrow A_2}\otimes\hat U_{\rm CNOT}^{B_1\rightarrow B_2}$.\\
%Two local controlled-NOT gates are performed on the pairs. \\
({\bf B4}) Measure the target pair ($A_2,B_2$). \\
({\bf B5}) If the measurement result is either $\ket{00}$ or $\ket{11}$, perform
a $\hat{\sigma}_y^{A_1}$ rotation; otherwise discard pair ($A_1,B_1$).

These steps are applied to a whole ensemble and result in halving the number of pairs and yielding a new ensemble with bipartite states $\hat{\rho}'$. The fidelity of the pairs in the 
new ensemble $F'=\bra{\Psi^{-}} \hat{\rho}' \ket{\Psi^{-}}$ is larger than the fidelity of the pairs in the processed ensemble $F=\bra{\Psi^{-}} \hat{\rho} \ket{\Psi^{-}}$ provided that
initially $\bra{\Psi^-} \hat{\rho} \ket{\Psi^-}>\frac{1}{2}$. Now, these steps are repeated from the beginning and this iteration leads to the purification of $\ket{\Psi^-}$. 
The requirement for the initial state $F=\bra{\Psi^-} \hat{\rho} \ket{\Psi^-}>\frac{1}{2}$ can be overcome by a 
certain filtering operation, aimed to exploit entanglement in a different way \cite{Horodecki}.

Let us briefly recapitulate step ({\bf B1}) due to its use in our subsequent discussions. 
A general bipartite state can be converted to the Werner state  
\begin{eqnarray}
	\hat{\rho}_W(F) &=& F \ket{\Psi^{-}}\bra{\Psi^{-}} + \frac{1-F}{3} \ket{\Psi^{+}}\bra{\Psi^{+}} \nonumber \\
	&+& \frac{1-F}{3}  \ket{\Phi^{-}}\bra{\Phi^{-}}+ \frac{1-F}{3} \ket{\Phi^{+}}\bra{\Phi^{+}},  
%	\nonumber \\ F&=&\bra{\Psi^{-}} \hat{\rho} \ket{\Psi^{-}},
	\label{wernerstate}
\end{eqnarray}
with the help of a linear projection \cite{Werner}, called also as the twirling operation. 
It has also been shown that $12$ local random unitary operations from the $SU(2)$ group, 
are necessary and sufficient to bring any two-qubit state $\hat{\rho}$ into a Werner
state \cite{Bennett3}; four operations are needed to bring $\hat \rho$ into a state $\hat \rho_{\rm BD}$ which is 
diagonal in the Bell basis and in turn three more operations transform $\hat \rho_{\rm BD}$ into a Werner state 
$\hat\rho_W$ (we will omit the dependence on $F$ when no ambiguity arises).
This can be written as
\begin{eqnarray}
  \hat \rho_W= \frac{1}{3} \sum_{j=1}^3B^\dagger_j\hat\rho_{\rm BD}\hat B_j,\quad
  \hat \rho_{\rm BD}=\frac{1}{4}\sum_{j=1}^4 \hat B_j^\dagger \hat B_j^\dagger\hat \rho \hat B_j \hat B_j, \nonumber \\
  \label{twirlingop}
\end{eqnarray}
where we have used the $4$ unitary transformations
\begin{align}
  \hat B_j&=\hat b_j^A\otimes \hat b_j^B,\quad
  \hat b_1^\ell=\frac{\hat{\mathbb{I}}^\ell+i\hat\sigma_x^\ell}{\sqrt2},\quad
  \hat b_2^\ell=\frac{\hat{\mathbb{I}}^\ell-i\hat\sigma_y^\ell}{\sqrt2},\quad
  \nonumber\\
  \hat b_3^\ell&=\ketbra{1}{1}^\ell+i\ketbra{0}{0}^\ell,\quad \hat b_4^\ell=\hat{\mathbb{I}}^\ell,
  \quad \ell\in\{A,B\}.
  \label{Bs}
\end{align}
which have been expressed in terms of the local unitary transformations $\hat b_j$ acting on a single qubit, the Pauli spin operators $\hat\sigma_x$ and $\hat\sigma_y$
and the identity map $\hat{\mathbb{I}}$.  
All three states have the same fidelity with respect to $\ket{\Psi^-}$, i.e
\begin{equation}
  F=
  \bra{\Psi^-}\hat \rho\ket{\Psi^-}=
  \bra{\Psi^-}\hat \rho_{\rm BD}\ket{\Psi^-}=
  \bra{\Psi^-}\hat \rho_W\ket{\Psi^-}.
  \label{}
\end{equation}

In our scheme we consider that each qubit pair flies through cavities on side $A$ and $B$ and after two sequential interaction of the qubits with the single-mode fields
two postselective field measurements are performed. This method generates two probabilistic two-qubit quantum operations on the two pairs on side $A$ and $B$ as shown in Sec. \ref{Model}. 
These quantum operations replace the controlled-NOT operations used in the original purification procedure. 
Our alternative version of the protocol ({\bf aB}) requires the following four steps:

\noindent
({\bf aB1}) We assume that the every spatially separated pair is entangled and is brought in the Werner state by 
local random unitary operations. This is equivalent to ({\bf B1}). We denote a four-qubit state by $\hbrho$.
Therefore, the four-qubit input state reads
\begin{equation}
\hbrho=\hat{\rho}_W^{A_1,B_1} \otimes \hat\rho_{W}^{A_2,B_2}
\label{Werner1}
\end{equation}
with the Werner state defined in Eq. \eqref{wernerstate}.\\ 
({\bf aB2}) We apply now the two-qubit quantum operations which results in the state
\begin{equation}
\hbrho'=
%\frac{\hat{M}^{A_1,A_2} \hat{M}^{B_1,B_2} \hbrho \hat{M}^{A_1,A_2} \hat{M}^{B_1,B_2} }{\mathrm{Tr} 
%\left\{\left(\hat{M}^{A_1,A_2} \hat{M}^{B_1,B_2}\right)^2 \hbrho\right\}}
\frac{\hat Q \hbrho \hat Q^\dagger }{\mathrm{Tr} 
\left\{\hat Q^\dagger \hat Q \hbrho\right\}}
,\quad \hat Q=\hat{M}^{A_1,A_2}\otimes \hat{M}^{B_1,B_2},
\label{map}
\end{equation}
where $\hat M^{\ell_1,\ell_2}$ is the matrix $\hat M$ in Eq. \eqref{Qgate} acting on qubits $\ell_1$ and $\ell_2$, with $\ell\in\{A,B\}$. 
The success probability to obtain this state
is given by the normalization factor
\begin{equation}
 \Tr{\hat Q^\dagger \hat Q \hbrho}=\frac{5 - 4 F + 8 F^2}{18}.
 \label{exprob}
\end{equation} 
({\bf aB3}) One of the pairs is now locally measured, for instance ($A_2,B_2$). 
We remark that in our scheme the two qubit pairs can be 
treated symmetrically.
\\
({\bf aB4}) Depending on the four possible measurement events we use the following strategy: in the cases when both qubits are in $\ket{0}$ or $\ket{1}$ we
apply a local unitary transformation $\hat \sigma_x^{A_1}$ on the unmeasured pair; otherwise do nothing. This step is fundamentally different
from ({\bf B5}) because there are no inadequate measurement results and we do not have to discard the unmeasured pair. 

\begin{figure}[t!]
 \includegraphics[width=.45\textwidth]{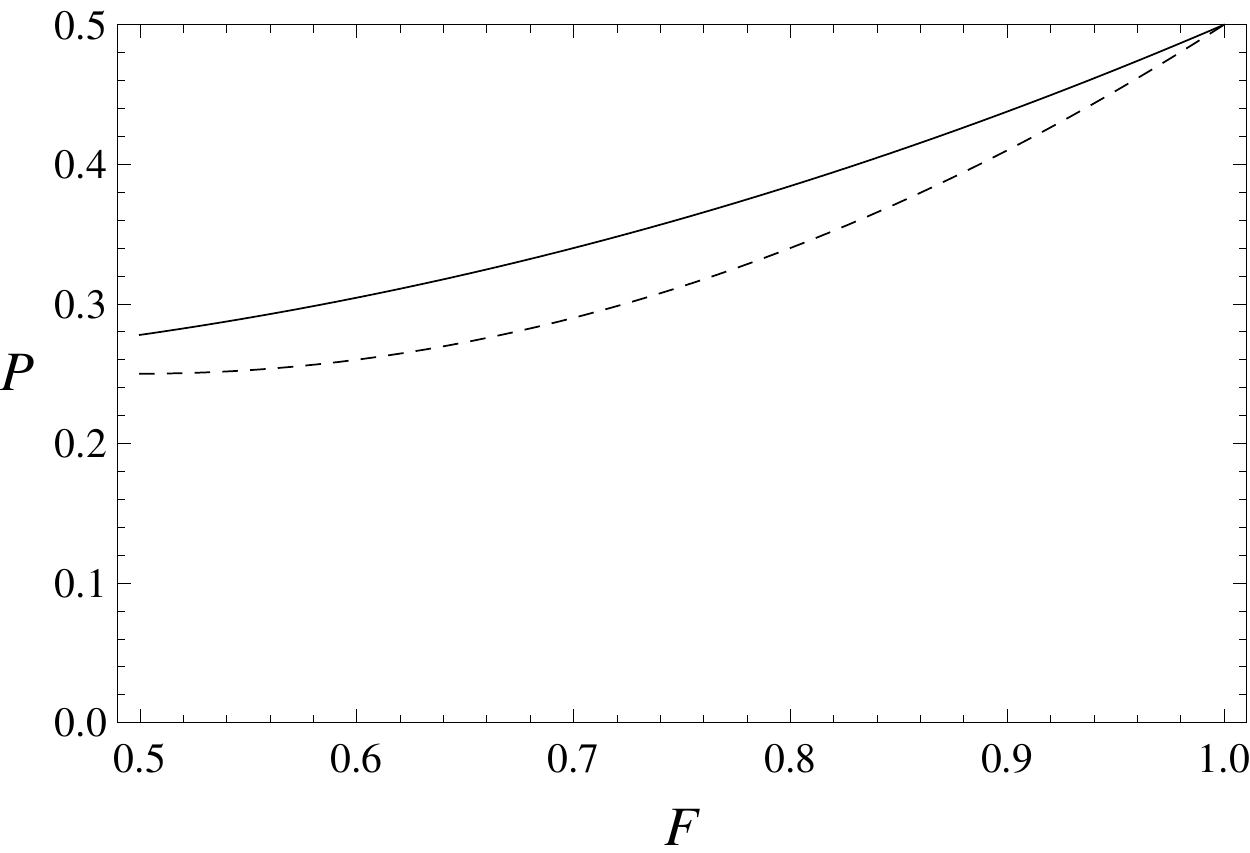}
 \includegraphics[width=.45\textwidth]{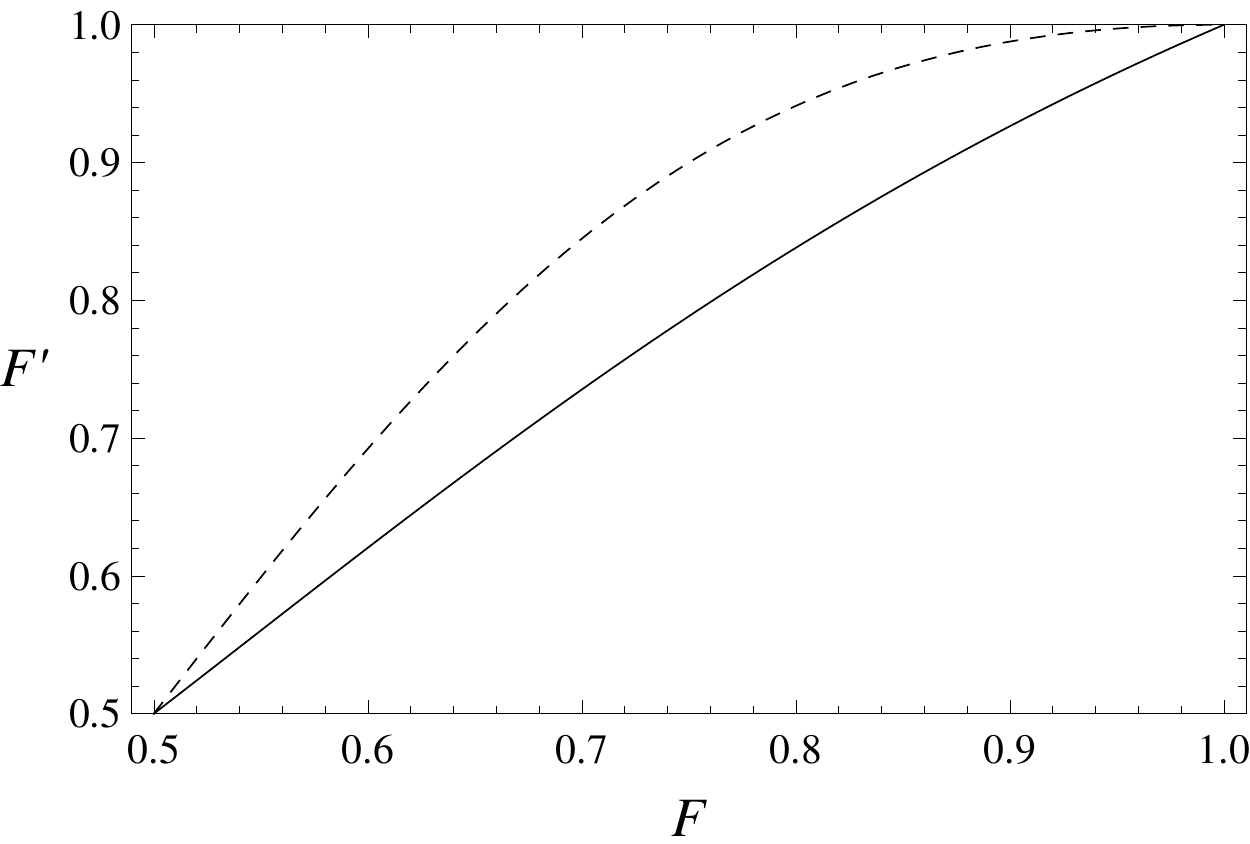}
  \caption{Top panel: The success probability of the entanglement purification protocols. Bottom panel: The achieved new fidelities after a successfully applied protocol.
  Both figures are shown for the initial state in Eq. \eqref{nWerner}.  The plots show in full line the results of protocol ({\bf aB})  
  and in dashed line the results of protocol ({\bf aD}). 
  }
  \label{probfid}
\end{figure}

It is interesting to note that the success probability of the protocol is determined by step ({\bf aB2}) compared with the original scheme of Ref. \cite{Bennett1} where
the selective measurement on the qubit pairs in step ({\bf B5}) specifies this probability. Provided that we are successful in the photonic postselection we generate a bipartite state 
$ \hat \rho'$ with fidelity
\begin{equation}
 \bra{\Psi^{-}} \hat{\rho}' \ket{\Psi^{-}}=F'=\frac{1 - 2 F + 10 F^2}{5 - 4 F + 8 F^2}.
 \label{newfid}
\end{equation}
This is exactly the same equation obtained in Ref. \cite{Bennett1} and our scheme has a success probability $P=(5 - 4 F + 8 F^2)/18$. The dependency on $F$ for both
the new fidelity $F'$ and the success probability $P$ are shown in Fig. \ref{probfid}.

Let us consider now an input four-qubit state with different fidelities
\begin{equation}
\hbrho=\hat{\rho}_W^{A_1,B_1}(F_1) \otimes \hat\rho_{W}^{A_2,B_2}(F_2).
\label{Werner2}
\end{equation}
Applying the purification protocol we obtain the following fidelity 
\begin{equation}
 F'=\frac{1-F_1-F_2+10F_1F_2}{5-2F_1-2F_2+8F_1F_2}
\end{equation}
with success probability
\begin{equation}
 P=\frac{5-2F_1-2F_2+8F_1F_2}{18}.
\end{equation}
If one chooses  $F_1=0.4$ and $F_2=0.75$, then the purification protocol generates a bipartite state
with fidelity $F'=0.558$. In general this means that the ensemble of pairs can have different fidelities and the only condition of a successful purification is 
that the average fidelity of the ensemble is larger than $0.5$.

{\it Scheme 2.}  
Now we turn our attention to the method in Ref. \cite{Deutsch} which is conceptually similar to Ref. \cite{Bennett1} and operates not on Werner states but on 
states diagonal in the Bell basis
\begin{align}
 \hat{\rho}_{\rm BD}(F,F_1,F_2,F_3) &= F \ket{\Psi^{-}}\bra{\Psi^{-}} 
	+F_1  \ket{\Phi^{-}}\bra{\Phi^{-}} 
 \nonumber \\&+ 
	F_2 \ket{\Phi^{+}}\bra{\Phi^{+}}
 + F_3 \ket{\Psi^{+}}\bra{\Psi^{+}} 
\end{align}
with $F+F_1+F_2+F_3=1$. In the case when we start initially with an arbitrary state, then a twirling operation with four unitary operators (see Eq. \eqref{twirlingop})
is required in order to bring this state in a Bell diagonal form. We remark that this scheme purifies state $\ket{\Phi^+}$, therefore increasing
the value of $F_2$.
The protocol for two qubit pairs can be summarized in four steps:\\ 
({\bf D1}) Apply the unitary operation 
$\hat b_1^{\dagger A_1}\otimes\hat b_1^{\dagger A_2}\otimes\hat b_1^{B_1}\otimes\hat b_1^{B_2}$, see Eq. \eqref{Bs} .\\ 
%at $A$:
%\begin{equation}
% \ket{0}\rightarrow \frac{1}{\sqrt{2}} (\ket{0}-i\ket{1}),\,\,\, \ket{1}\rightarrow \frac{1}{\sqrt{2}} (\ket{1}-i\ket{0}).
%\end{equation}
%Perform the inverse operation at $B$
%\begin{equation}
% \ket{0}\rightarrow \frac{1}{\sqrt{2}} (\ket{0}+i\ket{1}),\,\,\, \ket{1}\rightarrow \frac{1}{\sqrt{2}} (\ket{1}+i\ket{0}).
%\end{equation}
({\bf D2}) Perform the bilateral operation  $\hat U_{\rm CNOT}^{A_1\rightarrow A_2}\otimes\hat U_{\rm CNOT}^{B_1\rightarrow B_2}$.\\
({\bf D3}) Measure the target pair ($A_2,B_2$). \\
%(D3) The target qubit pair is measured. \\
({\bf D4}) If the measurement result is either $\ket{00}$ or $\ket{11}$ then
the unmeasured pair is kept; otherwise is discarded.

In our alternative scheme ({\bf aD}) we purify again with respect to $\ket{\Psi^-}$. Provided that an ensemble of Bell diagonal 
states is generated among the flying qubits we proceed with the following four steps:\\
%(D1) This step is neglected.\\
({\bf aD1}) To the four-qubit input state 
\begin{equation}
\hbrho=\hat{\rho}_{\rm BD}^{A_1,B_1}(F,F_1,F_2,F_3) \otimes \hat\rho_{\rm BD}^{A_2,B_2}(F,F_1,F_2,F_3).
\end{equation}
we directly apply the two-qubit quantum operation of Eq. \eqref{map}.
We obtain a $\hbrho'$  with success probability
$(F+F_1)^2/2+(F_2+F_3)^2/2$.\\
({\bf aD2}) The same as ({\bf aB3}).\\
({\bf aD3}) The same as ({\bf aB4}).\\
({\bf aD4}) Apply the rotation $\hat b_3^{A_1}\otimes \hat b_3^{B_1}$.  

Provided that we are successful we obtain the following Bell diagonal state:
\begin{equation}
 \hat{\rho}'_{\rm BD}\left(\frac{F^2+F_1^2}{D},\frac{2F_2F_3}{D},\frac{2FF_1}{D},\frac{F_2^2+F_3^2}{D}\right)
\end{equation}
with $D=(F+F_1)^2+(F_2+F_3)^2$. Our step ({\bf aD4}) is analog to the step ({\bf D1}) and flips the Bell states $\ket{\Phi^\pm}$ while leaving $\ket{\Psi^\pm}$
invariant. This map redistributes the fidelities in order to obtain a purification by iteration. This result is similar to the one obtained in Ref. \cite{Deutsch}, with the only
difference that protocol {\bf aD} purifies with respect to the state $\ket{\Psi^-}$ and  protocol {\bf D} purifies with respect to the state $\ket{\Phi^{+}}$.

An interesting feature arises when one  considers the following special Bell diagonal state 
\begin{equation}
 \hat{\rho}_\Psi(F)=F \ket{\Psi^-}\bra{\Psi^-}+(1-F) \ket{\Psi^+}\bra{\Psi^+}
 \label{nWerner}
\end{equation}
which is naturally generated in proposals for hybrid quantum repeaters \cite{vanLoock1, Bernad1,Gonta2}. 
The four-qubit state reads
\begin{equation}
 \hbrho= \hat{\rho}^{A_1,B_1}_\Psi(F) \otimes  \hat{\rho}^{A_2,B_2}_\Psi(F).
\end{equation}
The step ({\bf aD4}) in our protocol is actually not required for this type of initial states, as we never populate the states
$\ket{\Phi^\pm}$. However, this step will prove to be of crucial importance when applying a noisy version
of the operation $\hat M$ of Eq. \eqref{map} such as our proposed cavity-QED version in Eq. \eqref{W2}.

Thus our protocol ({\bf aD}) yields the bipartite state
\begin{eqnarray}
\hat\rho_\Psi\left(\frac{F^2}{1-2F+2F^2}\right),
\label{newfid2}
\end{eqnarray}
with success probability
\begin{equation}
 P=\frac{1-2F+2F^2}{2}.
 \label{probaD}
\end{equation}

%%%%%%%%%%%%%%%%%%%%%%%%%%%%%%%%%%%%
%%%%%%%%%%%%%%%%%%%%%%%%%%%%%%%%%%%%%%%

In the case of different input fidelities
\begin{equation}
 \hbrho= \hat{\rho}^{A_1,B_1}_\Psi(F_1) \otimes  \hat{\rho}^{A_2,B_2}_\Psi(F_2)
\end{equation}
we obtain the bipartite state
\begin{eqnarray}
\hat\rho_\Psi\left(\frac{F_1F_2}{(1-F_1)(1-F_2)+F_1F_2}\right)
\end{eqnarray}
with success probability
\begin{equation}
 P=\frac{1-F_1-F_2+2F_1F_2}{2}.
\end{equation}

%%%%%%%%%%%%%%%%%%%%%%%%%%%%%%%%%%%%
\begin{figure}[t]
 \includegraphics[width=.45\textwidth]{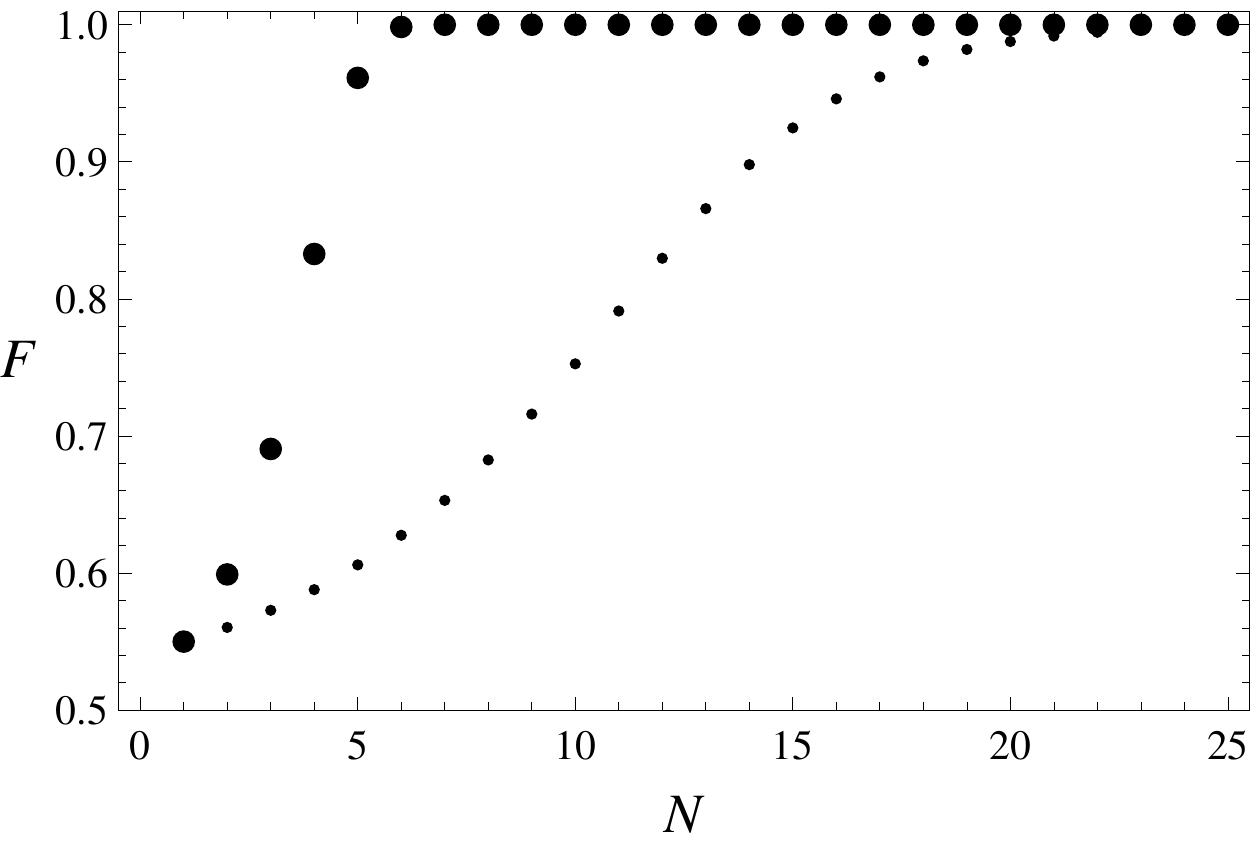}
 \includegraphics[width=.46\textwidth]{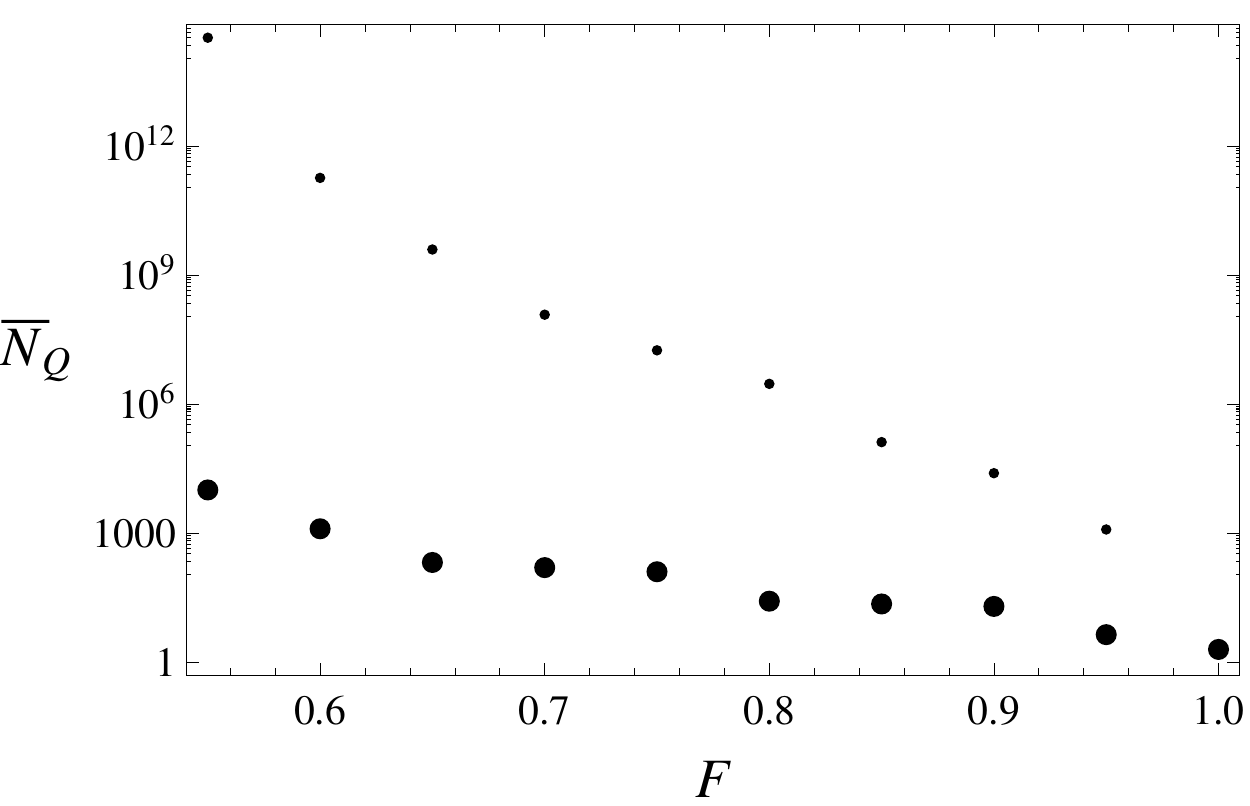}
  \caption{Top panel: The fidelity of the states achieved after $N$ successful iterations and based on Eqs. \eqref{newfid} (normal dots) and \eqref{newfid2} (thicker dots).
  Bottom panel: The average number of qubit pairs $\overline{N}_{Q}$ required to reach the final fidelity of $0.99$ as a function of the initial fidelity $F$. 
  As an inital condition we considered a special Bell diagonal state given in Eq. \eqref{nWerner}. The plots show that 
  the purification scheme ({\bf aD}) (thicker dots) outperforms the purification scheme ({\bf aB}) (normal dots).}
  \label{itdependence}
\end{figure}
%%%%%%%%%%%%%%%%%%%%%%%%%%%%%%%%%%%%%%%

In Fig. \ref{probfid} we compare the fidelity and success probability obtained from both of our protocols ({\bf aB}) 
in full line and ({\bf aD}) in dashed line for initial states of the form of Eq. \eqref{nWerner}. 
The function $F'(F)$ in dashed line shows a more concave shape than the full line counterpart. This means that less
iterations are required in order to attain almost unit fidelity as shown in the top panel of Fig. \ref{itdependence}.
The success probability for the protocol ({\bf aD}) is slightly lower than that of ({\bf aB}) as shown in the bottom panel of Fig. 
\ref{probfid}. However, the number of the average qubit pairs needed for the purification is more sensitive to the number of
iterations required as it is shown in the bottom panel of Fig. \ref{itdependence}. These numbers were obtaind with the help of 
the fidelity dependent probabilities in Eqs. \eqref{exprob} and \eqref{probaD}. Thus the protocol ({\bf aD}) is more efficient than protocol 
({\bf aB}) and it is also the most robust against noisy implementations as it is demonstrated in the subsequent section.

\section{Effects of a one-atom maser on the purification scheme}
\label{loss}

%%%%%%%%%%%%%%%%%%%%%%%%%%%%%%%%%%%%
\begin{figure}[t]
 \includegraphics[width=.45\textwidth]{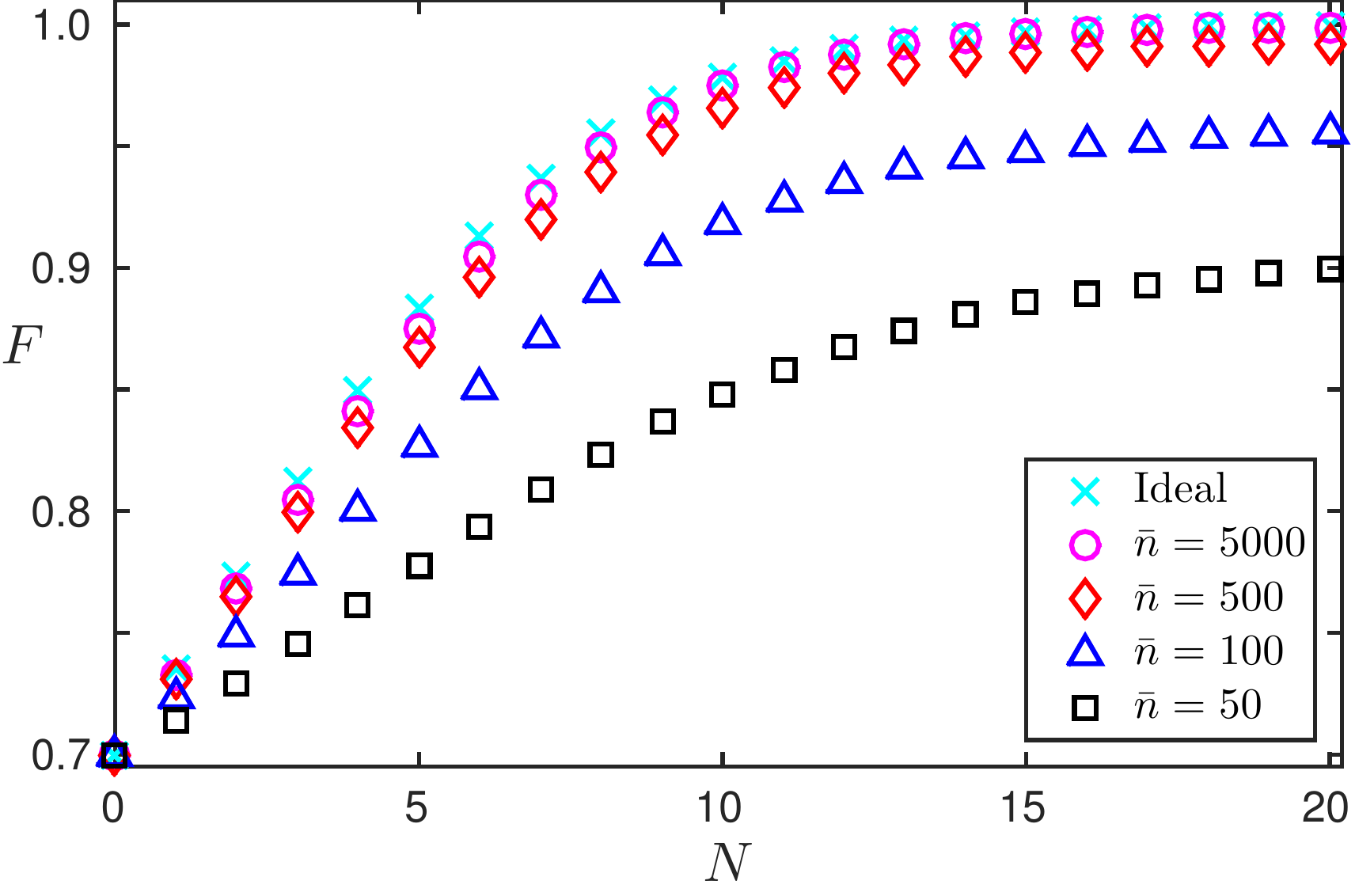}
 \includegraphics[width=.45\textwidth]{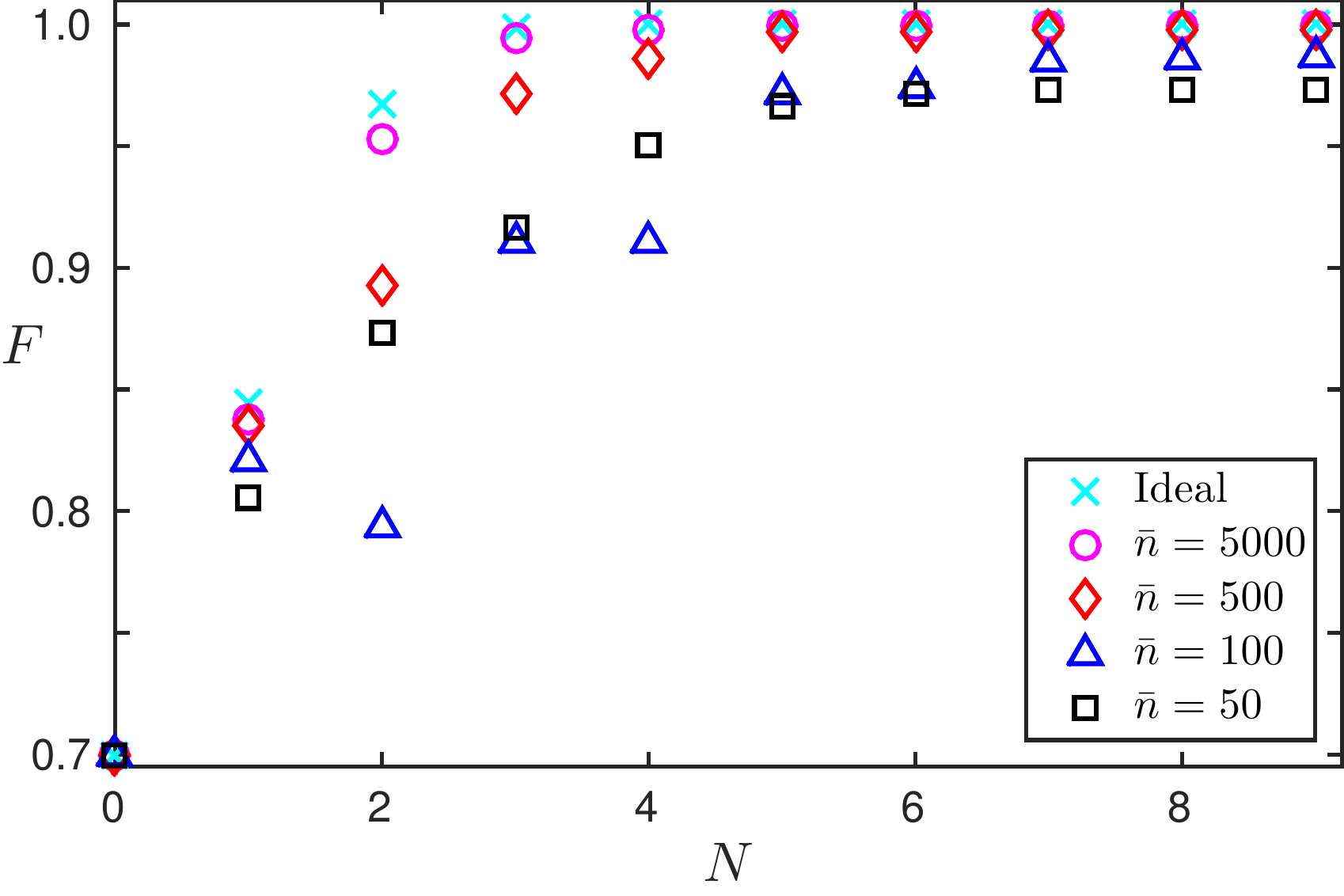}
 \caption{
 The achieved new fidelities with respect to the Bell state $\ket{\Psi^-}$ after several successful iterations and for 
 different values of the mean photon number $\bar n$. Top panel: Protocol ({\bf aB}).
 Bottom panel: Protocol ({\bf aD}). Both figures are considered for the same initial state of Eq. \eqref{nWerner} with fidelity $F=0.7$. 
 We employ the two-qubit quantum operation $\hat W_2(0.5,\bar n)$
 in Eq. \eqref{W2} with different values of the mean photon number $\bar n$ and use non-identical interaction times for system $A_1$ and $A_2$, i.e., $\tau_{A_2}/\tau_{A_1}=1.1$.
 }
  \label{fiditer}
\end{figure}
%%%%%%%%%%%%%%%%%%%%%%%%%%%%%%%%%%%%%%%

In this section we analyze the physical boundaries of our model proposed in Sec. \ref{Model} in the application
of the entanglement purification protocols of Sec.  \ref{Purify}. 
We consider the two-qubit quantum operation $\hat W_2(p,\bar n)$ of Eq. \eqref{W2} as an approximation of
the  entangling two-qubit quantum operation $\hat M$ of Eq. \eqref{Qgate} which is the core element in our protocol.
The value of the quadrature $p$ is obtained by a balanced-homodyne measurement of the field. 
The approximation becomes more accurate with increasing values of the mean photon number $\bar n$ of the initial coherent state
and provided that the interaction time $\tau$ for each atom 
fulfills  the condition of Eq. \eqref{taucondition}.  
In the experimental setting of S. Haroche \cite{Hagley,Nogues} the interaction time $\tau$ is not equal for each atom.
However, it can be shown that for any two atoms $A_1$, $A_2$ the interaction times fulfill 
the inequality $|\tau_{A_1}/\tau_{A_2}-1|\leqslant 0.01$. 
This is achieved  by Doppler-selective optical pumping 
techniques  \cite{Hagley,Nogues} that significantly reduce the width of the velocity distribution,  which directly affects the matter-field interaction times. 

In Fig. \ref{fiditer}, we have plotted 
the fidelities as a function of the iterations $N$ for the protocols presented in Sec. \ref{Purify}, 
where we employ $\hat W_2(0.5,\bar n)$ in place of $\hat M$. Additionally we have taken into account that the first (second) atom 
interacts for a time $2/g$ ($2.2/g$)  with the field in order to demonstrate the stability with respect to deviations in the interaction times. 
The result for protocol ({\bf aB}) is in the top panel  and for ({\bf aD}) in the bottom panel
for an initial states of the form of Eq. \eqref{nWerner}. 
 With this initial states the step ({\bf aD4}) is unnecessary 
when using the perfect two-qubit quantum operation $\hat M$.
In contrast this step ({\bf aD4}) plays a crucial role with the operation
$\hat W_2$ and initial states of \eqref{nWerner}.
We ran simulations (not shown here) without step ({\bf aD4}) and found that the fidelity
drops  after few iterations due to the noisy quantum operation $\hat W_2(p,\bar n)$ 
that populates the other Bell states $\ket{\Phi^\pm}$. We have considered an ensemble of qubit pairs with moderate input fidelity $F=0.7$. We see that the protocol
({\bf aD}) attains high fidelities quite rapidly in $N=5$ iterations outperforming protocol ({\bf aB}) in regard to the average number of qubit pairs needed for the 
purification. Taking into account the success probability of protocol ({\bf aD}), one would require an average number of 
$2600$ qubit pairs to obtain a final fidelity of $0.999999$. This simple analysis suggests that the core mechanism of our purification scheme is feasible taking into account
typical experiments where $35700$ atomic samples are sent through a cavity \cite{Haroche2}. 
For the interaction times we have chosen realistic parameters based on Ref. \cite{Haroche2} that 
reports an interaction time of $24\, \mu s$ ($6\, mm$ waist / $250\, m s^{-1}$), atoms separated by $70\, \mu s$ 
time intervals, and coupling strengths of $g=2 \pi \times 51$ kHz.

Now we turn our attention to the effects of photonic losses in our protocol described by
the cavity damping rate $\kappa$ and the spontaneous emission $\gamma$ of the 
atoms.  State of the art microwave cavities present very low values of $\kappa$ \cite{Haroche2}. However, 
our protocol requires the cavity field to leak in order to implement a balanced homodyne photodetection. 
For this reason it would be more favorable to use a cavity with a Q-factor in between the current technology
and previous realizations that present ratios of roughly $g/\kappa=60$ \cite{Haroche}. In such  case
$3\, ms$ is enough to empty the cavity ($e^{-3}\approx 0.05$) and measuring a single
quadrature takes $1\, \mu s$ \cite{Hansen} or $5.5\, ns$ \cite{Cooper}. 
Considering these time scales and the fact that the Rydberg atoms used in the S. Haroche experiments have a 
ratio of $g/\gamma=3000$,  it is to be expected that atomic spontaneous decay does not play a major 
destructive effect in one step of our protocol. Nevertheless, it could play a role  during the iteration procedure and therefore the qubits coherency 
must be kept in order that the purification procedure works. We do not elaborate on this here, but  merely 
estimate that if the realization of one iteration is dominated by cavity leakage of time $3\, ms$, 
then protocols above $N=10$ iterations are sensitive to the spontaneous decay of the atoms.

In the presence of losses the ideal two-qubit quantum operation $\hat{\rho} \rightarrow \hat{M} \hat{\rho} \hat{M}$ has to be replaced  by a more general quantum operation
$\hat{\rho} \rightarrow \mathcal{E}\hat{\rho}$ which depends on $\kappa$ and $\gamma$. 
In the following we investigate numerically the effect of this general quantum operation
on the entanglement purification protocol. We consider a Markovian description in which the 
evolution of an initial density matrix $\hat\varrho_0$, describing both atoms and the cavity, is given by
\begin{equation}
  \hat\varrho(\tau,\tau_f)=\cV(\tau,\tau_f)\hat\varrho_0,\quad \cV(\tau,\tau_f)= 
  e^{\cL^{A_2}\tau}e^{\cL_{f}\tau_f}
  e^{\cL^{A_1}\tau}.
\label{lossmodel}      
\end{equation}
The evolution operator $\cV(\tau,\tau_f)$ is generated by the Liouvillians  
\begin{equation}
  \cL^\ell\hat\varrho=i\left[\hat \varrho,\hat H^\ell\right]+
  \cL_f
  \hat\varrho
  ,
\quad
\cL_f=\cL_{\rm HO}+\cL_{\rm at}^{A_1}+\cL_{\rm at}^{A_2},
%\nonumber\\
  %(
  %\cL_{\rm HO}
  %+\cL_{\rm at}^A+\cL_{\rm at}^B)
  \label{Liouvillian1}
\end{equation}
with  $\ell\in\{A_1,A_2\}$,  that have been written in terms of the dissipators 
\begin{align}
  \cL_{\rm HO}\hat\varrho&=
  \tfrac{1}{2}\kappa(n_T+1)
  \left(2\hat a \hat\varrho\hat a^\dagger-\hat a^\dagger \hat a\hat\varrho-\hat\varrho\hat a^\dagger \hat a\right)
  \nonumber\\
  &\qquad\,\,\,\,+
  \tfrac{1}{2}\kappa n_T
  \left(2\hat a^\dagger \hat\varrho\hat a-\hat a\hat a^\dagger \hat\varrho-\hat\varrho\hat a\hat a^\dagger\right),
  \nonumber\\
  \cL_{\rm at}^\ell\hat\varrho&=\tfrac{1}{2}\gamma
  \left(2\hat\sigma_-^\ell\hat\varrho\hat\sigma_+^\ell
  -\hat\sigma_+^\ell\hat\sigma_-^\ell\hat\varrho-\hat\varrho\sigma_+^\ell\hat\sigma_-^\ell \right)
  \label{}
\end{align}
which describe the losses of the cavity and spontaneous emission of the atoms respectively. 
The evolution operator $\cV(\tau,\tau_f)$ reflects the fact that at all times the dissipation
mechanisms are active in the system, while the interaction only happens first for time $\tau$
between cavity and atom $A_1$ and  for the same amount of time $\tau$ between cavity and atom $A_2$.
In between the interactions there is a time of free evolution $\tau_f$ where only dissipation effects
take place. 
In typical microwave experiments, the average number
of thermal photons $ n_T$ is equal to $0.05$ towards which the field evolves with rate $\kappa$ (see Ref. \cite{Haroche2}). 
 The initial condition is taken to be the same as in Eq. \eqref{initial}. 

In order to efficiently compute the dynamics for high photon numbers, we evaluate the quantum operation 
\begin{align}
\cE(\tau,\tau_f,p)\hat\rho
&=\bra p\left(\cV(\tau,\tau_f)\hat \rho\ketbra{\alpha}{\alpha}\right)\ket p
\nonumber\\
&=\sum_{i,j,k,l=0}^3\cE_{k,l,i,j} \rho_{i,j}\ketbra{\varphi_k}{\varphi_l},
\label{Channel}
\end{align}
where $\ket{\varphi_i}\in\{\ket{00},\ket{01},\ket{10},\ket{11}\}$ and $\rho_{i,j}=\bra{\varphi_i}\hat \rho \ket{\varphi_j}$. One can find that the entries of $\cE=\cE(\tau,\tau_f,p)$ are given by
\begin{align}
\cE_{k,l,i,j}=\Tr{\ketbra{\varphi_l}{\varphi_k}\otimes\ketbra{p}{p} \cV(\tau) \ketbra{\alpha}{\alpha}\otimes \ketbra{\varphi_i}{\varphi_j}}.
\label{ChannelEntries}
\end{align}
The quantum operation in Eq. \eqref{Channel} is the noisy analog to the quantum operation in Eq. \eqref{W2}.
Ideally, when $\bar n \gg 1$ and the decay constants tend to zero then $\cE\hat\varrho\to\hat M \hat\varrho \hat M$. In the same way as 
the quantum operation $\hat M \cdot \hat M$, $\cE \cdot$ also  does not preserve the trace.   
For the quantum purification protocol, the noisy analog to Eq. \eqref{map}  takes the following form
\begin{equation}
\hbrho'= \frac{\cE^{A_1,A_2}\cE^{B_1,B_2}\hbrho}{\Tr{\cE^{A_1,A_2}\cE^{B_1,B_2}\hbrho}}.
\label{}
\end{equation}

%%%%%%%%%%%%%%%%%%%%%%%%%%%%%%%%%%%%
\begin{figure}[t!]
 \includegraphics[width=.45\textwidth]{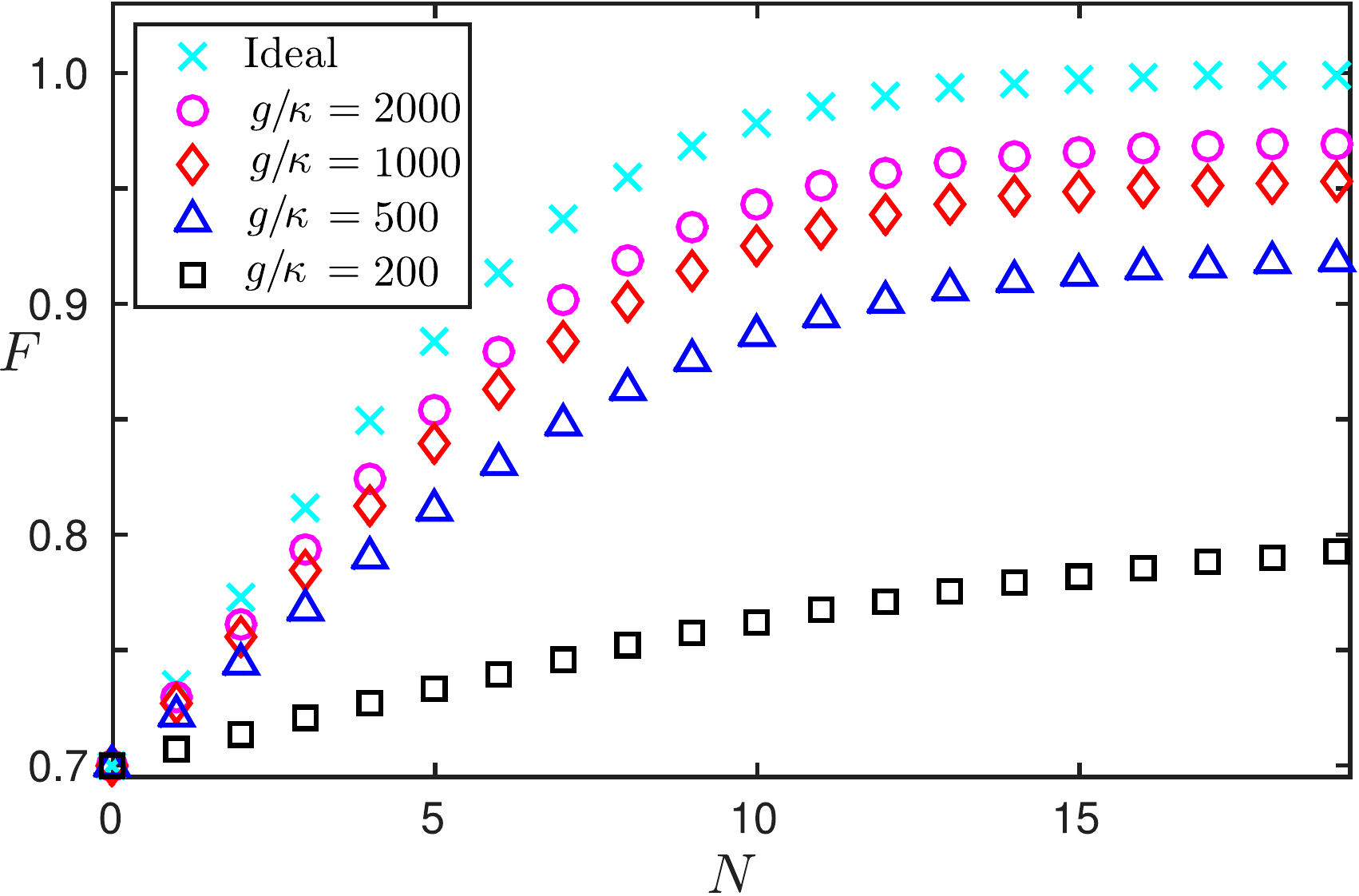}
 \includegraphics[width=.45\textwidth]{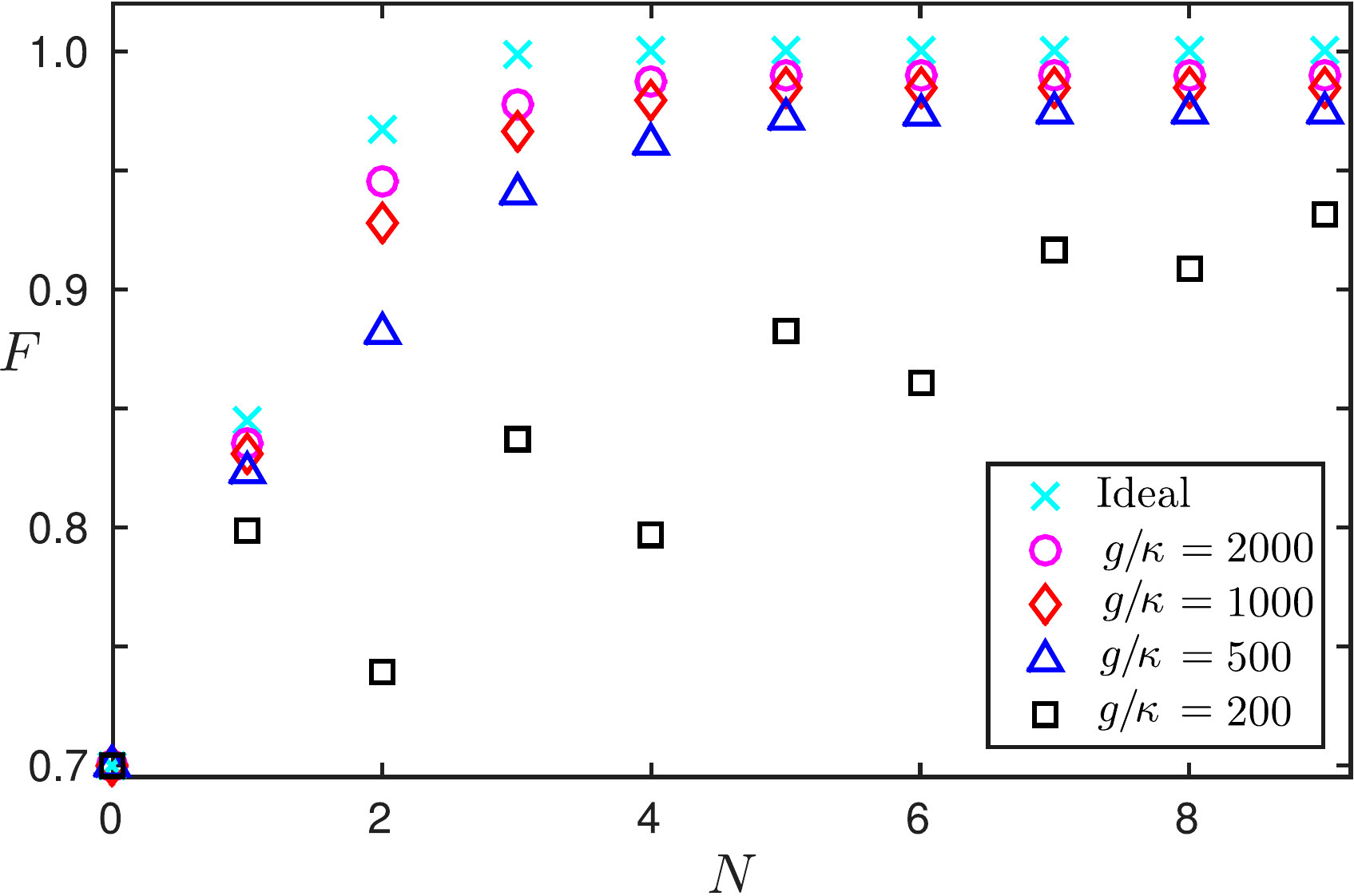}
 \caption{
 The achieved fidelities with respect to the Bell state $\ket{\Psi^-}$ after several successful iterations and for different values of the cavity decay rate $\kappa$.
 Top panel: Protocol ({\bf aB}). Bottom panel: Protocol ({\bf aD}). Both figures are considered for the same initial state of Eq. \eqref{nWerner} with fidelity $F=0.7$. 
 We employ the quantum operation $\cE(2/g,3/g,0.15)$ in Eq. \eqref{Channel} for a mean photon number $\bar n =500$. The crosses 
 show the ideal two-qubit quantum operation $\hat{\rho} \rightarrow \hat M \hat{\rho} \hat M $. Spontaneous decay rate was set to 
 $g/\gamma=3000$ and we considered an average thermal photon number $n_T=0.1$.}
  \label{fiditerloss}
\end{figure}
%%%%%%%%%%%%%%%%%%%%%%%%%%%%%%%%%%%%%%%

In Fig. \ref{fiditerloss} we plot the achieved fidelity after several successful iterations 
using different values of the decay constant $\kappa$ and an average thermal photon number $n_T=0.1$. 
We have numerically evaluated the quantum operation $\cE \cdot$ as indicated in Eq. \eqref{ChannelEntries}. 
We have considered a chopped Hilbert space with $N_F=\lfloor \bar n+ 4\sqrt{\bar n} \rfloor$ \cite{floor} 
Fock states and constructed an $4N_F^2\times 4N_F^2$ matrix describing the Liouvillians in \eqref{Liouvillian1}. 
We have chosen the value of the quadrature $p$ to be $0.15$, on which the field state is projected. Numerical investigations show that an increase in the 
absolute value of $p$ implies a slightly decreased performance in the 
purification protocols. This is due to the lossy dynamics which brings closer the outer field contributions and thus distorting the boundaries 
of the central peak (see Fig. \ref{fig:QA} for the ideal case). Therefore, for quadrature values being farthest from the origin in the interval $[-2,2]$ 
we obtain more noisy versions of the ideal two-qubit quantum operation $\hat{\rho} \rightarrow \hat M \hat{\rho} \hat M$. Provided that we use the parameters
of the experimental setup in Ref. \cite{Haroche2} the quadrature measurements around the central peak always generate a high fidelity two-qubit quantum operation
in regard to the ideal one.  The time of the free evolution is set to be larger 
than the interaction time in order to simulate almost the same conditions which are present in experimental scenarios. 
It can be noticed that protocol ({\bf aD}) is more robust against the effects of losses
and surprisingly $N=5$ iterations are required to achieve its maximum fidelity. 
In this case, the step ({\bf aD4}) plays a crucial role in the stabilization of the protocol.

\section{Conclusions}
\label{conclusions}

We have discussed implementations of entanglement purification protocols in the context of a hybrid quantum repeater. Our scheme is based
on the one-atom maser, thus making our proposal a good experimental candidate. It has been demonstrated that a probabilistic two-qubit quantum operation
can be realized with the help of ancillary multiphoton states. The two qubits fly sequentially through a
single-mode cavity, initially prepared in a coherent state, and interact with the radiation field. The emerged field state is measured by a balanced homodyne photodetection. 
We have shown that for resonant matter-field interactions and large values of the  
mean photon number, the two-qubit quantum operation in Eq. \eqref{Qgate} can be implemented
with high fidelity. This is based on the fact that for interaction times characterizing the collapse phenomena in the Jaynes-Cummings-Paul model the field contribution
correlated with this quantum operation can be perfectly distinguished from the other field contributions correlated with other components of the two-qubit state. 
We have shown that the obtained probabilistic two-qubit quantum operation can replace the controlled-NOT gate in standard purification protocols \cite{Bennett1, Deutsch}. 
This approach have resulted in two alternative purification protocols, called in the main text ({\bf aB}) and ({\bf aD}), 
which are conceptually similar to their standard versions. These new protocols discard qubit pairs due to unsuccessful photonic postselection, but in the case of qubit measurements
all the unmeasured qubit pairs are kept and only a measurement dependent unitary rotation is performed on them. 
We have compared these protocols for initial states which are in a special Bell diagonal
form and they are generated in the proposals for hybrid quantum repeaters.
 
Finally, we have investigated the role of the losses in our proposed scheme. We have taken into account the damping rate of the cavity and the spontaneous decay of the qubits. We have 
conducted numerical investigations which show that our scheme is sensitive to the cavity damping rate in the sense that high fidelities $F>0.95$ can be achieved but
never unit fidelity. These numerics were based on parameters taken from real experimental setups. There is also a trade-off between good and bad cavities because high-$Q$ cavities
enhance the fidelity of the two-qubit quantum operation, but on the other hand the leakage of the field which has to be measured takes a longer time, thus increasing the chance of a
spontaneous decay in the qubits. In general, we have found that protocol ({\bf aD}) which does not employ the twirling operation is more efficient
than protocol ({\bf aB}) by means of the average number of qubit pairs needed for obtaining high fidelity Bell states.
Furthermore, protocol ({\bf aD}) can correct errors in the implementation
of the two-qubit quantum operation.   

In view of recent developments in quantum communication and quantum state engineering this work might offer interesting perspectives. The results clearly show
the limitations of a purification protocol in a hybrid quantum repeater based on multiphoton states, but on the positive side the proposed scheme has a high repetition rate.
The proposed scheme can be already implemented in a one-atom maser setup. However, 
other implementations  
may include condensed-matter qubits which are coupled to
single-mode radiation fields \cite{Sun}, trapped ions inside a cavity \cite{Casabone}, and  neutral atoms coherently transported into an optical resonator \cite{Reimann}. 

\begin{acknowledgments}
This work is supported by the BMBF project Q.com.
\end{acknowledgments}

\appendix
\section{The states of the field}
\label{App}
In this appendix we present the unnormalized field states %$g_{ij}(\tau)$ 
which appear in equation~\eqref{psi}. They are defined by
\begin{align}
	\ket{g_{00}(\tau)} &= \sum_{n=0}^{\infty} \expal \frac{\alpha^n}{\sqrt{n!}} \Big [ c_{00} \cos\left(\Omega_{n-1} \tau \right) \cos\left(\Omega_{n-1} \tau \right) \ket{n}  \nonumber \\
	&-  i c_{10} \sin \left(\Omega_n \tau \right) \cos\left(\Omega_{n} \tau \right) \ket{n+1}  \nonumber \\
	&-  i c_{01} \cos\left(\Omega_{n-1} \tau \right) \sin \left(\Omega_n \tau \right) \ket{n+1}  \nonumber \\
	&-  c_{11} \sin \left(\Omega_n \tau \right)\sin \left(\Omega_{n+1} \tau \right) \ket{n+2}  \Big] \; ,
\end{align}
\begin{align}
	\ket{g_{01}(\tau)} &= \sum_{n=0}^{\infty} \expal \frac{\alpha^n}{\sqrt{n!}} \Big[ c_{01} \cos\left(\Omega_{n-1} \tau \right) \cos\left(\Omega_{n} \tau \right) \ket{n} \nonumber \\ 
	&-  i c_{00} \frac{\alpha}{\sqrt{n+1}} \cos\left(\Omega_{n} \tau \right) \sin\left(\Omega_{n} \tau \right) \ket{n}  \nonumber \\ 
	&-  i c_{11} \sin\left(\Omega_{n} \tau \right) \cos\left(\Omega_{n+1} \tau \right) \ket{n+1}  \nonumber \\ 
	&-  c_{10} \sin\left(\Omega_{n} \tau \right) \sin\left(\Omega_{n} \tau \right) \ket{n} \Big] \; ,
\end{align}
\begin{align}
	\ket{g_{10}(\tau)} &= \sum_{n=0}^{\infty} \expal \frac{\alpha^n}{\sqrt{n!}} \Big[ c_{10} \cos\left(\Omega_{n} \tau \right) \cos\left(\Omega_{n-1} \tau \right) \ket{n} \nonumber \\
	&-  i c_{11} \cos\left(\Omega_{n} \tau \right) \sin\left(\Omega_{n} \tau \right) \ket{n+1}  \nonumber \\
	&-  i c_{00} \frac{\alpha}{\sqrt{n+1}} \sin\left(\Omega_{n} \tau \right) \cos\left(\Omega_{n-1} \tau \right) \ket{n}  \nonumber \\
	&-  c_{01}  \sin\left(\Omega_{n-1} \tau \right)  \sin\left(\Omega_{n-1} \tau \right) \ket{n} \Big] \; ,
\end{align}
\begin{align}
	\ket{g_{11}(\tau)} &= \sum_{n=0}^{\infty} \expal \frac{\alpha^n}{\sqrt{n!}} \Big[ c_{11} \cos\left(\Omega_{n} \tau \right) \cos\left(\Omega_{n} \tau \right) \ket{n} \nonumber \\ 
	&-  i \frac{\alpha}{\sqrt{n+1}} c_{10} \cos\left(\Omega_{n+1} \tau \right) \sin\left(\Omega_{n} \tau \right) \ket{n} \nonumber \\
	&-  i \frac{\alpha}{\sqrt{n+1}} c_{01} \sin\left(\Omega_{n} \tau \right) \cos\left(\Omega_{n} \tau \right) \ket{n} \nonumber \\
	&-  \frac{\alpha^2}{\sqrt{(n+1)(n+2)}} c_{00} \sin\left(\Omega_{n+1} \tau \right)  \sin\left(\Omega_{n} \tau \right)  \ket{n} \Big] \; ,
\end{align}
where $\Omega_n = g \sqrt{n+1}$.


\begin{thebibliography}{99}
\bibitem{floor}  $\lfloor x\rfloor$ is  largest integer less than or equal to $x$.
\end{thebibliography}
\end{document}